\documentclass[a4paper,fleqn]{cas-sc}

\usepackage{xcolor}
\definecolor{darkgreen}{rgb}{0.0, 0.5, 0.0}

\usepackage{amsmath}
\usepackage{amssymb}
\usepackage[numbers]{natbib}
\usepackage{subcaption}
\usepackage{caption}
\usepackage{booktabs}
\usepackage{multirow}
\usepackage[section]{placeins}
\usepackage[switch]{lineno}

\graphicspath{{Figs/}}

\ExplSyntaxOn
\cs_set:Npn \__first_footerline: { }
\ExplSyntaxOff

\begin{document}
\let\WriteBookmarks\relax
\def\floatpagepagefraction{1}
\def\textpagefraction{.001}


\shorttitle{Neural TO for AIP minimization in ship structures}
\shortauthors{S. Yan et al.}

\title[mode=title]{Neural topology optimization of ship structures under propulsion machinery vibrations}

\author[1]{Shengyu Yan}[orcid=0000-0002-9077-9649]

\author[1]{Muhammad Muztahidul Hakim~Zareer}

\author[1,2]{Jasmin Jelovica}[orcid=0000-0002-8396-941X]
\cormark[1]

\ead{jjelovica@mech.ubc.ca}

\affiliation[1]{%
  organization={Department of Mechanical Engineering,
    The University of British Columbia},
  addressline={6250 Applied Science Ln},
  city={Vancouver},
  postcode={V6T 1Z4},
  country={Canada}%
}

\affiliation[2]{%
  organization={Department of Civil Engineering,
    The University of British Columbia},
  addressline={6250 Applied Science Ln},
  city={Vancouver},
  postcode={V6T 1Z4},
  country={Canada}%
}

\cortext[cor1]{Corresponding author}

\begin{abstract}
Ship structural vibrations contribute to noise, fatigue, and equipment damage. Dynamic-compliance topology optimization can produce pathological designs near resonance. This study extends neural-reparameterized topology optimization using a convolutional Kolmogorov--Arnold network (KATO) to forced-vibration design by incorporating active input power (AIP) as the objective. Applications comprise a 100~Hz engine-supporting deck panel and an 18~Hz thruster foundation frame. Helmholtz PDE filtering, Heaviside projection, and three projected deck fields control feature sizes and define a manufacturing-tolerance envelope. Across both deck families, all eight layouts reduce AIP relative to family-matched size-optimized references. After finite-depth extrusion, all eight outperform their references in AIP and static compliance.  Corrected 5--500~Hz modal sweeps characterize off-design response. For the unrestricted, manufacturing-aware, and stress-aware frame variants, KATO matches GCMMA in AIP within 0.5~dB and yields 22--36$\times$ lower static compliance after matched-volume binary re-analysis. The convolutional generator provides a structural prior and implicit regularization, producing near-binary, connected load paths that remain effective after thresholding. In a matched near-resonant 300~Hz case, both methods reduce initial AIP by more than 32~dB; KATO preserves one connected component, achieves 59$\times$ lower binary static compliance, and reduces the maximum AIP over 1--500~Hz by 2.7~dB. On the shared accelerated FEA backend, KATO runs 6.4--10.4$\times$ faster than GCMMA for the implemented stress-aware formulations, while the corrected 200-mode model evaluates deck frequency points 990--2125$\times$ faster than the full-order solution. These results establish neural AIP-driven topology optimization as an efficient approach for designing connected, feature-size-controlled ship structures with improved forced-vibration performance.
\end{abstract}


\maketitle

\section{Introduction}\label{sec:intro}

Structural vibrations have several unwanted consequences, including reduced fatigue life, passenger and crew discomfort, and noise. In
marine vessels, machinery-induced vibration can contribute to underwater
radiated noise (URN) \cite{ross1987mechanics}. The current IMO guidelines
identify URN from shipping as a source of adverse impacts on marine life and
set out measures for its reduction \cite{imo2024urn}. Vibration generated by
propulsion and auxiliary machinery is transmitted through foundations and deck
panels and can radiate from the hull skin
\cite{junger1986sound, lin2009study}. Reducing the vibrational energy injected
into these structural paths is a primary engineering objective.

Stiffened panels, resilient mounts, and constrained-layer damping treatments
address ship vibration effectively in many configurations but depend on the
designer's judgment regarding stiffener placement and sizing. Such designs
satisfy static strength and global stiffness requirements yet are rarely tuned
for dynamic vibration transmission at specific excitation frequencies. Topology
optimization (TO) offers a systematic alternative: given a design domain,
boundary conditions, and loading, TO distributes material to minimize
structural objectives while respecting volume and manufacturing constraints
\cite{bendsoe2003topology, sigmund2013topology}.

Since the homogenization-based formulation of Bends{\o}e and Kikuchi
\cite{bendsoe1988generating}, density-based methods have become the dominant
paradigm. The SIMP method \cite{bendsoe1989optimal, sigmund200199}
parameterizes each finite element with a continuous density variable and
penalizes intermediate densities to drive the design toward a binary (0--1)
distribution. Alternative representations include level-set methods
\cite{wang2003level, allaire2004structural}, evolutionary structural
optimization (ESO/BESO) \cite{xie1993simple, young19993d,
huang2010evolutionary}, and topological derivative methods
\cite{sokolowski1999topological, eschenauer1994bubble}; Sigmund and Maute
\cite{sigmund2013topology} provide a comparative review.
The computational cost of TO grows with problem size, particularly for 3D
domains and repeated dynamic solves. Large-scale TO has been enabled by
parallel computing \cite{aage2017giga}, and recent work addresses acceleration
strategies for industrial problems and compares major TO formulations on industrial-like 3D benchmarks \cite{mukherjee_accelerating_2021,
yago2022topology}. Numerical instabilities---checkerboard patterns, mesh
dependence, local minima---remain persistent challenges
\cite{sigmund1998numerical}, addressed by density filtering
\cite{bourdin2001filters, svanberg2013density} and Heaviside projection
\cite{wang2011projection}.

Extending TO to dynamic structural problems has followed two principal
directions. The first manages the eigenvalue spectrum: maximizing fundamental
frequencies, widening frequency gaps, or shifting natural frequencies away
from excitation bands \cite{diaz1992solutions, ma1993structural,
pedersen2000maximization, du2007topological}. These formulations suit
free-vibration problems but do not directly address forced-vibration amplitude
or power flow. The second direction targets forced-vibration response under
harmonic or broadband loading. Dynamic compliance, i.e. the work done by external
forces on the displacement field, was historically used as the objective
\cite{olhoff2016generalized, olhoff2014intro}, but it exhibits sign changes,
singularities, and non-monotonic behavior near resonance and antiresonance
frequencies that destabilize gradient-based optimizers
\cite{silva2019critical}. These pathologies are particularly severe in
single-frequency optimization, where the optimizer may exploit
antiresonance-induced sign flips rather than genuinely reducing vibration.
Rong et al.\ \cite{rong2000dynamic_constraints} addressed forced-vibration
optimization through dynamic response constraints, Takezawa et al.\
\cite{takezawa2016complex} proposed complex dynamic compliance for damped
structures, and Saurabh et al.\ \cite{saurabh2024robust} developed robust
formulations for transient response. In structural acoustics, Du and Olhoff
\cite{du2007minimization} minimized radiated sound power from vibrating
bi-material structures, Neofytou et al.\ \cite{neofytou2025automatic}
demonstrated automatic-differentiation-based 3D acoustic-structural TO, and
Hu et al.\ \cite{HU2023116843} explored coupled acoustic--mechanical
optimization of porous structures.

An alternative objective that avoids the sign ambiguity of dynamic compliance is
active input power (AIP), proposed by Silva et al.\ \cite{silva2020topology}
following their critical analysis of dynamic compliance
\cite{silva2019critical}. AIP is the time-averaged real power injected by
harmonic forces into the structural velocity field. It is non-negative for
passive structures, avoids sign ambiguity, and represents the total mechanical
energy input that must be either dissipated or radiated as acoustic power.
Silva et al.\ \cite{silva2021shape} further extended AIP-based optimization to
shape and position-preserving objectives. To date, however, AIP-based TO has
been applied mainly to academic benchmarks. Its application to large engineering problems including ship structural
components has not been explored.

Topology optimization has been applied to ship structures. Cui et al. \cite{cui2015structural} combined knowledge-based engineering, surrogate modelling, multi-objective optimization, and a level-set method to reduce container-ship tank weight and redesign brackets. Daifuku et al. \cite{daifuku2016design} used frequency-response finite-element analysis to develop engine-room reinforcements that suppress generator-induced vibration. Jia et al. \cite{jia2019design} optimized a trimaran bulkhead under seven load cases, obtaining layouts similar to conventional stiffener arrangements. Kendibilir and Kefal \cite{kendibilir2023enhanced} applied peridynamics-based topology optimization to a cracked dredger cross-section under hogging and sagging loads, producing stiffer web-frame concepts. Yilmaz and Konal \cite{yilmaz2025topology} converted optimization results into a manufacturable welded hatch cover, reporting a 2.49-t reduction and projected operational benefits. Cao et al. \cite{cao2026topological} optimized an unmanned sailboat wing sail, hull, keel, and rudder and verified improvements in weight, stability, and wave resistance using coupled multi-physics simulations. When topology optimization is used for large-scale ship structures, computational cost becomes a major bottleneck due to fine discretizations, numerous load cases, and coupled multidisciplinary analyses. Although surrogate modelling has been explored in a limited context, its systematic integration into large-scale, multidisciplinary ship-structure topology optimization remains insufficiently investigated.

Separately, machine learning has been applied to TO through trained neural
predictors of optimal layouts \cite{li2019non, oh2019deep, chi2021universal}
and broader ML-assisted workflows \cite{shin2023topology,
mukherjee_accelerating_2021}. These data-driven approaches accelerate
inference but depend on training datasets and may generalize poorly outside
their distributions. Neural reparameterization takes a different approach:
an \emph{untrained} network parameterizes the density field during
optimization, with the architecture imposing an implicit spatial prior toward
smooth, connected topologies. Introduced by Hoyer et al.\
\cite{hoyer2019neural} and drawing on the deep image prior
\cite{ulyanov2018deep}, this paradigm has been developed through TOuNN
\cite{chandrasekhar2021tounn}, TONR \cite{zhang2021tonr},
physics-informed frameworks \cite{jeong_physics-informed_2023}, and
training-dataset-free approaches \cite{liu2026topology}. Gradient-free
variants have also been proposed \cite{kus2024gradient, kato2023tackling}.

Our prior work, KATO \cite{yan2025kato}, introduced convolutional
Kolmogorov--Arnold network (cKAN) generators \cite{bodner2024convolutional,
liu2024kan} to neural reparameterized TO, demonstrating improved structural
connectivity and stress performance relative to standard CNN generators.
KATOsuper \cite{yansuper2025kato} extended the framework to volumetric 3D
optimization and introduced sensitivity-consistent Fourier neural operator
(SC-FNO) surrogates. The present work applies the KATO-cKAN architecture to dynamic forced-vibration problems, targeting AIP minimization in an engine-supporting ship deck and a thruster-supporting frame.

This paper makes the following contributions:
\begin{enumerate}
  \item \textbf{A dynamic TO for ship structures based on active input power.}
    The forced-vibration pipeline combines AIP with
    static-compliance regularization, Helmholtz filtering and three-field
    projection for feature-size control, stress-aware variants, and
    binary post-evaluation. This extends the KATO generator
    framework from static objectives to frequency-domain forced-vibration
    design.
  \item \textbf{Systematic optimization comparison under realistic loading.}
    Topology-optimized deck layouts are benchmarked against size-optimized
    stiffened single-skin and X-core double-skin panels under machinery loading
    representative of
    marine propulsion systems.
  \item \textbf{A neural-versus-conventional optimizer comparison on a thruster foundation structure.} The KATO generator
    is compared against the globally convergent method of moving asymptotes
    (GCMMA) under a common mixed AIP/static base
    objective.
  \item \textbf{Analysis of stress and feature-size control.} The
    trade-off between vibration suppression, structural compliance, stress
    $p$-norm, and local maximum stress is quantified across regularization
    strategies, including Helmholtz filtering and stress-aware penalization. The behavior of the neural parameterization is contrasted with a
    density-filtered conventional baseline.
\end{enumerate}

\section{Problem Formulation}\label{sec:formulation}

\subsection{Density-based topology optimization}

Following the SIMP density approach \cite{bendsoe2003topology}, a design domain
$\Omega$ is discretized into $N_e$ finite elements, each assigned a density
$\rho_i \in (0, 1]$. Stiffness follows the penalized power law:
\begin{equation}\label{eq:simp}
  E_i = E_{\min} + \rho_i^{p} (E_0 - E_{\min})
\end{equation}
where $E_0$ is the base material Young's modulus, $E_{\min} = 10^{-9} E_0$
prevents singularity, and $p > 1$ is the penalization exponent (ramped during
continuation).

Mass density follows a \emph{linear} interpolation rather than the penalized
stiffness law \cite{pedersen2000maximization}:
\begin{equation}\label{eq:mass_interp}
  \varrho_i = \varrho_{\min} + \rho_i \,(\varrho_0 - \varrho_{\min})
\end{equation}
where $\varrho_0$ and $\varrho_{\min}$ are the solid and void material
densities. This choice avoids the artificially high local frequencies produced
when mass is penalized more strongly than stiffness. It can instead introduce
low-frequency modes localized in weak, low-density regions. Such modes have
negligible participation under the applied loads; resonance locations are thus
identified from the load-specific frequency response rather than from the
smallest eigenvalue alone.

\subsection{Frequency-domain dynamic equation}

Under harmonic excitation at angular frequency $\omega$, the steady-state
displacement field $\mathbf{U}(\omega,\boldsymbol{\rho}) \in \mathbb{C}^{N_{\mathrm{dof}}}$
satisfies the frequency-domain equation of motion:
\begin{equation}\label{eq:freq_domain}
  \mathbf{S}(\omega, \boldsymbol{\rho})\,\mathbf{U}(\omega,\boldsymbol{\rho})
  = \mathbf{F}(\omega)
\end{equation}
Here, $\boldsymbol{\rho}$ denotes the element density vector,
$\mathbf{U}(\omega,\boldsymbol{\rho})$ and $\mathbf{F}(\omega)$ are the
complex displacement and force amplitude vectors, respectively, and
$\mathbf{S}(\omega,\boldsymbol{\rho})$ is the dynamic stiffness matrix. The
density dependence of $\mathbf{U}$ enters through $\mathbf{S}$, so the shorter
notation $\mathbf{U}(\omega)$ is used when no ambiguity arises. The matrix
$\mathbf{S}$ is defined as
\begin{equation}\label{eq:dynamic_stiffness}
  \mathbf{S}(\omega, \boldsymbol{\rho})
  = \mathbf{K}(\boldsymbol{\rho})
  - \omega^2 \mathbf{M}(\boldsymbol{\rho})
  + \mathrm{i}\,\omega\,\mathbf{C}(\boldsymbol{\rho})
\end{equation}
where $\mathbf{K}(\boldsymbol{\rho})$, $\mathbf{M}(\boldsymbol{\rho})$, and
$\mathbf{C}(\boldsymbol{\rho})$ are the global stiffness, mass, and damping
matrices, respectively, and $\mathrm{i}=\sqrt{-1}$.
The broadband calculations use stiffness-proportional structural damping with
a constant loss factor $\eta_s$:
\begin{equation}\label{eq:structural_damping}
  \mathbf{S}(\omega,\boldsymbol{\rho})
  = \mathbf{K}(\boldsymbol{\rho})-
    \omega^2\mathbf{M}(\boldsymbol{\rho})+
    \mathrm{i}\,\eta_s\mathbf{K}(\boldsymbol{\rho}) .
\end{equation}
We set $\eta_s=0.02$, corresponding to $\zeta\simeq\eta_s/2=1\%$ for light
damping, and use this value throughout the reported analyses. At the design
frequency $\omega_d$, this operator is
equivalent to stiffness-proportional viscous damping with
$\beta=\eta_s/\omega_d$. Across a broadband sweep, the constant loss factor
avoids the increase $\zeta=\beta\omega/2$ produced by a fixed Rayleigh
coefficient.

Because $\mathbf{S}$ is complex-valued and non-Hermitian in general, the system
in Eq.~\eqref{eq:freq_domain} is recast as an equivalent $2N_{\mathrm{free}}
\times 2N_{\mathrm{free}}$ real block system by splitting $\mathbf{U} = \mathbf{u}_r
+ \mathrm{i}\,\mathbf{u}_i$:
\begin{equation}\label{eq:real_block}
  \begin{bmatrix} \mathbf{A} & -\mathbf{B} \\ \mathbf{B} & \mathbf{A} \end{bmatrix}
  \begin{bmatrix} \mathbf{u}_r \\ \mathbf{u}_i \end{bmatrix}
  = \begin{bmatrix} \mathbf{f}_r \\ \mathbf{f}_i \end{bmatrix}
\end{equation}
where $\mathbf{A} = \mathbf{K} - \omega^2\mathbf{M}$ and
$\mathbf{B} = \omega(\alpha\mathbf{M} + \beta\mathbf{K})+
\eta_s\mathbf{K}$ are the real and imaginary parts of $\mathbf{S}$ restricted
to free DOFs, respectively; $\alpha$ and $\beta$ are the mass- and
stiffness-proportional Rayleigh damping coefficients. The viscous and structural-loss contributions can be used
separately; the reported broadband sweeps use $\alpha=\beta=0$ and
$\eta_s=0.02$.
This formulation enables the use of high-performance sparse LU factorization
(PARDISO \cite{schenk2004solving}) for the frequency-domain solve.

\subsection{Active input power objective}

The instantaneous mechanical power delivered by a harmonic force
$\mathbf{f}(t) = \mathrm{Re}[\mathbf{F}\,\mathrm{e}^{\mathrm{i}\omega t}]$
to the structural velocity field
$\dot{\mathbf{u}}(t) = \mathrm{Re}[\mathrm{i}\omega\mathbf{U}\,\mathrm{e}^{\mathrm{i}\omega t}]$
has a time-averaged value:
\begin{equation}\label{eq:aip_definition}
  \Pi_{\mathrm{in}}(\omega)
  = \frac{1}{2}\,\mathrm{Re}\!\left[\mathbf{F}^H \dot{\mathbf{U}}\right]
\end{equation}
where superscript $H$ denotes the Hermitian (conjugate transpose).
For optimization, $\dot{\mathbf{U}}=\mathrm{i}\omega\mathbf{U}$ is substituted
into Eq.~\eqref{eq:aip_definition}. For a passive structure the active input
power is non-negative ($\Pi_{\mathrm{in}} \geq 0$) and free of the sign ambiguity
that afflicts dynamic compliance near antiresonances
\cite{silva2020topology, olhoff2016generalized}. We minimize its magnitude as a safeguard against roundoff-scale negative values
near zero, while retaining the signed value for consistency checks. For the passive-system solutions considered here, this safeguard does not alter the positive AIP
away from numerical zero crossings.
The reported decibel value is
\begin{equation}\label{eq:aip_db}
  L_{\Pi}(\omega) = 10\log_{10}\!\left(
    \frac{\left|\Pi_{\mathrm{in}}(\omega)\right|}{\Pi_0}
  \right),
\end{equation}
where $\Pi_0=10^{-12}$~W is the standard reference power for a power level. AIP
measures the structure-borne power injected by the prescribed harmonic load; it
is computed \emph{in vacuo} and is not radiated acoustic power. Its absolute
level changes with the load amplitude, damping model, and reference power, so
only designs evaluated under the same finite-element model and load case are
compared directly.

\subsection{Mixed objective and stress-aware regularization}

In the present optimizations, AIP minimization without static-compliance regularization can produce weakly connected or hinge-like structures. Near
a resonance or a narrow load path, small changes in topology can reduce dynamic
stiffness and disconnect material paths more severely than they change total
mass. A mixed objective combines AIP with static
compliance $C$ \cite{sigmund200199} to regularize structural integrity:
\begin{equation}\label{eq:mixed}
  J = w_{\mathrm{AIP}}\,\frac{\Pi_{\mathrm{in}}(\omega)}{\Pi_{\mathrm{ref}}}
    + (1-w_{\mathrm{AIP}})\,\frac{C}{C_{\mathrm{ref}}}
\end{equation}
Here $\Pi_{\mathrm{ref}}$ and $C_{\mathrm{ref}}$ are reference values
computed on a uniform density field at the target volume fraction, and
$w_{\mathrm{AIP}} \in [0,1]$ controls the trade-off. The static compliance term
is evaluated from the static equilibrium problem
$\mathbf{K}(\boldsymbol{\rho})\mathbf{u}_s=\mathbf{F}_s$:
\begin{equation}\label{eq:static_compliance}
  C = \mathbf{F}_s^T\mathbf{u}_s .
\end{equation}
Setting $w_{\mathrm{AIP}}=1$ enables pure AIP minimization; $w_{\mathrm{AIP}}<1$
adds static-compliance regularization. The optimization objective uses the
normalized ratio $C/C_{\mathrm{ref}}$, whereas the results tables report the
unnormalized static compliance $C=\mathbf{F}_s^T\mathbf{u}_s$ in N\,m. Since
$C_{\mathrm{ref}}$, the static load, and the support conditions are case-specific,
these reported compliance values are compared only within a single case study.

The stress-aware variants use a common $p$-norm measure of element-center von
Mises stress recovered with the solid-material constitutive matrix. Static
stress uses $\mathbf{u}_s$. Dynamic stress is represented by a phase-independent measure formed from the real and
imaginary harmonic components and combined as
\begin{equation}\label{eq:vm_amplitude}
  \sigma_{\mathrm{vm}} = \sqrt{\sigma_{\mathrm{vm,r}}^2 + \sigma_{\mathrm{vm,i}}^2}.
\end{equation}
This combined measure provides a conservative upper bound on the cycle-maximum von Mises stress and is exact when the real and imaginary stress tensors are proportional. It is used consistently in all dynamic-stress comparisons.
For objective-based stress regularization, the mixed objective $J$ gains an
additional term,
\begin{equation}\label{eq:stress_obj}
  J_{\sigma} = J + w_{\sigma}\,\frac{\tilde{\sigma}_{p}}{\sigma_{\mathrm{ref}}}
\end{equation}
where $w_\sigma$ is a weight and $\sigma_{\mathrm{ref}}$ is a normalization
reference. Alternatively, the same measure can enter an inequality constraint,
$\tilde{\sigma}_p/\sigma_{\mathrm{lim}}-1\le 0$. The aggregated stress is
evaluated over solid-like material using a weighted $p$-norm:
\begin{equation}\label{eq:pnorm_stress}
  \tilde{\sigma}_{p} = \left(
    \frac{\sum_i w_i \sigma_{\mathrm{vm},i}^{p}}
         {\sum_i w_i + \epsilon}
  \right)^{1/p}
\end{equation}
where $w_i$ is a solid-region weight and $\epsilon$ is a small numerical
stabilizer. A hard solid mask sets $w_i=1$ when
$\rho_i\ge\rho_{\mathrm{thr}}=0.5$ and $w_i=0$ otherwise, so only material
elements enter the average. A smooth mask replaces this step with a sigmoid
weight during continuous optimization. The two choices change only the numerical
aggregation of element stresses; the von Mises recovery is the same. All
stress-aware optimization runs and binary post-evaluation metrics use $p=8$,
which gives a smooth approximation to the local maximum stress while avoiding
the stress singularity problem in void regions \cite{le2010stress,
bruggi2008mixed}. The deck and frame stress runs report solid-only aggregation on
present material; the dynamic case uses the combined von Mises measure of
Eq.~\eqref{eq:vm_amplitude}.

An optional grayness penalty $J_g = w_g \cdot \bar{g}$, where $\bar{g} =
(1/N_e) \sum_i 4\rho_i(1-\rho_i)$, is added for certain variants to
discourage intermediate densities during early continuation stages. The same
quantity, expressed as a percentage, is the standard \emph{measure of
non-discreteness}~\cite{sigmund2007morphology}
\begin{equation}\label{eq:mnd}
  M_{nd} = \frac{100}{N_e}\sum_{i} 4\rho_i(1-\rho_i)\;\;[\%],
\end{equation}
used to characterize the continuous field before thresholding: $M_{nd}=0\%$ for
a binary design and $M_{nd}=100\%$ for an all-gray ($\rho_i=0.5$) field.

Collecting the objective terms gives
\begin{equation}\label{eq:total_objective}
  J_{\mathrm{total}}
  = \underbrace{w_{\mathrm{AIP}}\frac{\Pi_{\mathrm{in}}(\omega)}{\Pi_{\mathrm{ref}}}
    + (1-w_{\mathrm{AIP}})\frac{C}{C_{\mathrm{ref}}}}_{\text{mixed AIP/static } J}
  + w_\sigma\frac{\tilde{\sigma}_p}{\sigma_{\mathrm{ref}}}
  + w_g\,\bar{g} ,
\end{equation}
where $w_g\,\bar{g}$ is an optional grayness penalty. KATO uses this objective with the volume-preserving density map and, when active, the additive stress term.
GCMMA uses the mixed objective $J$ with explicit volume and stress constraints, as detailed in Section~\ref{sec:optimization_strategies}.

\subsection{Adjoint sensitivity analysis}

Sensitivities of $\Pi_{\mathrm{in}}$ with respect to the density field are
computed via the adjoint method. The loss function is differentiated through
the sparse real-block solve (Eq.~\eqref{eq:real_block}) using a custom
\texttt{torch.autograd.Function} that stores the LU factorization for the
adjoint solve:
\begin{equation}\label{eq:adjoint}
  \frac{\mathrm{d}J}{\mathrm{d}\rho_i}
  = -\boldsymbol{\lambda}^T
    \frac{\partial \mathbf{S}_{\mathrm{blk}}}{\partial \rho_i}\,
    \mathbf{U}_{\mathrm{blk}}
  + \frac{\partial J_{\mathrm{expl}}}{\partial \rho_i}
\end{equation}
where $\boldsymbol{\lambda}$ satisfies the adjoint system
$\mathbf{S}_{\mathrm{blk}}^T \boldsymbol{\lambda}
= (\partial J / \partial \mathbf{U}_{\mathrm{blk}})^T$,
solved by back-substitution using the stored LU factors (no additional
factorization required). Here $\mathbf{S}_{\mathrm{blk}}$ is the real block
matrix in Eq.~\eqref{eq:real_block}, $\mathbf{U}_{\mathrm{blk}}=
[\mathbf{u}_r^T,\mathbf{u}_i^T]^T$, and $J_{\mathrm{expl}}$ denotes the explicit
density dependence of the objective. The resulting sensitivities are passed to
the selected optimizer.

\section{Methodology}\label{sec:methodology}

\subsection{Optimization strategies}\label{sec:optimization_strategies}

Two optimization strategies are considered. In the first strategy, KATO parameterizes the density
field with a cKAN generator and updates its parameters using the Adam gradient-based optimizer. The volume
fraction is imposed by the volume-preserving map described below, and the
stress-aware variants minimize $J_{\mathrm{total}}$ in
Eq.~\eqref{eq:total_objective}. As a density-based benchmark optimizer, GCMMA
instead updates the filtered element densities directly and enforces the
volume fraction through an inequality constraint in its constrained
optimization update. Its
stress-aware variants minimize the same mixed AIP/static objective $J$ while
enforcing $\tilde{\sigma}_p/\sigma_{\mathrm{lim}}-1\le 0$, with
$\sigma_{\mathrm{lim}}=26$~MPa. Both strategies use the same finite-element
responses, normalization references, and $p=8$ stress aggregation. The
comparison therefore changes the optimization parameterization and constraint
handling.

Table~\ref{tab:hyperparams} lists the objective weights. The Unrestricted and
Manufacturing-aware KATO variants set $w_\sigma=0$, while the stress-aware
variants activate the additive stress term. The GCMMA stress-aware variants use
the explicit stress constraint instead of $w_\sigma$.

\begin{table}[pos=h]
  \centering
  \caption{Objective weights and stress treatment used by KATO and GCMMA in
    the deck-panel and frame cases. The stress-aware KATO variants use the
    weighted stress term in Eq.~\eqref{eq:stress_obj}, whereas the stress-aware
    GCMMA variants impose an explicit 26~MPa stress constraint.}
  \label{tab:hyperparams}
  \footnotesize
  \begin{tabular}{@{}llccc@{}}
    \toprule
    Benchmark & Variant & $w_{\mathrm{AIP}}$ & $w_\sigma$ & $\sigma_{\mathrm{ref}}$ (MPa) \\
    \midrule
    Deck (double-skin)   & Unrestricted / Mfg.-aware & 0.9 & 0    & -- \\
                         & Stress-aware      & 0.9 & 0.05 & 35 \\
                         & Stress + Mfg.\    & 0.9 & 0.08 & 35 \\
    \midrule
    Deck (single-skin)   & Unrestricted / Mfg.-aware & 0.9 & 0    & -- \\
                         & Stress-aware      & 0.9 & 0.20 & 25 \\
                         & Stress + Mfg.\    & 0.9 & 0.25 & 25 \\
    \midrule
    Frame (KATO)         & Unrestricted / Mfg.-aware & 0.8 & 0    & -- \\
                         & Stress-aware      & 0.8 & 0.05 & 35 \\
                         & Stress + Mfg.\    & 0.8 & 0.08 & 35 \\
    \midrule
    Frame (GCMMA)        & Unrestricted / Mfg.-aware  & 0.8 & \multicolumn{2}{c}{volume constraint only} \\
                         & Stress-aware / Stress + Mfg.\ & 0.8 & \multicolumn{2}{c}{$+$ stress constraint $\sigma_{\mathrm{lim}}=26$~MPa} \\
    \bottomrule
  \end{tabular}
\end{table}

\subsection{KATO generator parameterization}\label{sec:generator}

Following KATO \cite{yan2025kato} and KATOsuper \cite{yansuper2025kato}, the
density field is parameterized by a convolutional Kolmogorov--Arnold network
(cKAN) generator $G_\theta$ with learnable B-spline activations. The generator
maps a trainable latent code $\mathbf{z}$ to a 2D logit field $\mathbf{l} \in
\mathbb{R}^{N_y \times N_x}$, which is then processed through the density
pipeline described below to produce a physical density field $\boldsymbol{\rho}
\in [0,1]^{N_e}$. Both the generator weights $\theta$ and the latent code
$\mathbf{z}$ are updated by Adam \cite{kingma2014adam}. A constrained sigmoid
enforces the prescribed mean density on the design field before filtering and
projection, and all sensitivities are obtained via PyTorch automatic
differentiation through the entire pipeline.

\subsection{Density pipeline and volume constraint}\label{sec:density_pipeline}

The physical density $\boldsymbol{\rho}$ is obtained from the raw logits
$\mathbf{l}$ through a sequence of operations:

\paragraph{Volume-preserving sigmoid.}
The mean density of the design field is enforced through a constrained sigmoid.
For the $N_d$ design elements, a scalar bias $b$ is found by bisection so that
the mean density equals the target $\bar{v}$:
\begin{equation}\label{eq:constrained_sigmoid}
  \rho_i = \sigma(l_i + b), \qquad
  \frac{1}{N_d}\sum_{i \in \mathcal{D}} \sigma(l_i + b) = \bar{v}
\end{equation}
where $\mathcal{D}$ denotes the designable elements and $\sigma$ is the logistic
sigmoid. Implicit differentiation through $b$ preserves the target volume in
the backward pass.

\paragraph{Density filtering.}
Density filtering smooths the density field, suppresses checkerboarding, and reduces mesh dependence.
The study uses either a \emph{neighborhood-weighted cone filter}
\cite{bourdin2001filters} with radius $r_{\min}$ applied via 2D convolution,
or a \emph{Helmholtz PDE filter}
\cite{lazarov2011filters}; the case-specific choices are given in Section~\ref{sec:cases}. Its finite-element form is:
\begin{equation}\label{eq:helmholtz_filter}
  (r^2 \mathbf{K}_f + \mathbf{M}_f)\,\mathbf{z}_f = \mathbf{T}_f\,\boldsymbol{\rho}
\end{equation}
where $\mathbf{K}_f$ and $\mathbf{M}_f$ are the filter stiffness and mass
matrices, $\mathbf{T}_f$ maps element densities to filter nodes, and
$r=r_{\min}/(2\sqrt{3})$. The filtered density field is
$\tilde{\boldsymbol{\rho}}=\mathbf{T}_f^{T}\mathbf{z}_f$. Homogeneous Neumann
boundary conditions are applied
($\nabla z_f\cdot\mathbf{n}=0$), where $\mathbf{n}$ is the outward boundary
normal. The physical radius suppresses
mesh-scale oscillations and introduces a mesh-independent smoothing length
scale.

\paragraph{Heaviside projection.}
After filtering, a smooth Heaviside approximation \cite{wang2011projection}
drives the density toward a binary 0--1 distribution:
\begin{equation}\label{eq:heaviside}
  \hat{\rho}_i = \frac{\tanh(\beta_H \mu) + \tanh\!\bigl(\beta_H(\tilde{\rho}_i - \mu)\bigr)}
                      {\tanh(\beta_H \mu) + \tanh\!\bigl(\beta_H(1 - \mu)\bigr)}
\end{equation}
where $\beta_H$ is the projection sharpness. For the single-field projection
used by the Helmholtz-filtered frame variants, $\mu$ is adjusted by bisection
to preserve the target volume.

\paragraph{Three-field manufacturing projection.}
For the Helmholtz-filtered deck variants, three projected density fields are
generated to assess manufacturing tolerance \cite{wang2011projection,
sigmund2013topology}. Three Heaviside
projections with distinct thresholds $\mu_d < \mu_i < \mu_e$ produce dilated ($\hat{\boldsymbol{\rho}}_d$, $\mu_d = 0.3$),
  intermediate ($\hat{\boldsymbol{\rho}}_i$, $\mu_i = 0.5$),
  and eroded ($\hat{\boldsymbol{\rho}}_e$, $\mu_e = 0.7$) density fields.
  The eroded field tests the survival of narrow solid connections, whereas the
  dilated field identifies whether small void features remain open under the
  prescribed geometric perturbation; the intermediate field defines the nominal
  topology.
  Together with the Helmholtz filter radius and Heaviside projection, the three fixed thresholds control solid
  and void feature sizes and define the geometric variation under erosion and
  dilation. In the final deck Helmholtz runs, both the AIP
  objective and the stress measure (when active) are evaluated on the
  intermediate field,
  while the eroded and dilated fields define its manufacturing-tolerance
  envelope.
Because the objective and stress measure use the intermediate field, the three projections provide feature-size control and geometric-tolerance characterization rather than optimization over all three realizations. The three deck thresholds remain fixed. The frame Helmholtz variants use a
single Heaviside projection with a bisection-adjusted threshold.
The density pipeline uses staged continuation: the SIMP penalty and projection
sharpness increase as the learning rate decreases. The schedules are fixed
within each matched comparison.

\subsection{Physical-unit FEA and differentiable implementation}\label{sec:efficient}

The physical-unit FEA uses SI geometry and steel properties. The 2D
optimizations use four-node plane-stress quadrilaterals, and the extruded 3D models
use eight-node solid elements. Stiffness and consistent mass matrices are
evaluated by standard Gauss integration with the physical element dimensions,
so modal frequencies and harmonic responses are obtained in Hz.

A single frequency-domain factorization per design serves the AIP and
dynamic-stress evaluation. After the real-block solve
(Eq.~\eqref{eq:real_block}) returns $(\mathbf{u}_r,\mathbf{u}_i)$, the active
input power is an $O(N_{\mathrm{dof}})$ inner product rather than a second solve,
\begin{equation}\label{eq:aip_discrete}
  \Pi_{\mathrm{in}}(\omega)
  = \tfrac{1}{2}\,\omega\left(
    \mathbf{f}_i^{T}\mathbf{u}_r - \mathbf{f}_r^{T}\mathbf{u}_i \right),
\end{equation}
which reduces to $-\tfrac{1}{2}\,\omega\,\mathbf{f}_r^{T}\mathbf{u}_i$ for a real
force vector. The element-center strain amplitudes are recovered from the same
$(\mathbf{u}_r,\mathbf{u}_i)$ by the vectorized operation
$\boldsymbol{\varepsilon}_e=\mathbf{B}_c\mathbf{U}_e$, where $\mathbf{B}_c$ is
the element-center strain--displacement matrix and $\mathbf{U}_e$ is the
complex element displacement vector. This is followed by the
solid-material constitutive relation and the combined von Mises definition in
Eq.~\eqref{eq:vm_amplitude}. Because the stress term is evaluated on the same
density field and frequency as the AIP term, the two share this forward
displacement, and the mixed AIP/stress objective needs no additional
frequency-domain factorization.

The forward LU factors are reused for the adjoint back-substitution, and the
fixed sparsity structure is cached across iterations. The sparse solve is
wrapped in a custom \texttt{torch.autograd.Function}; the remaining operations
use differentiable PyTorch tensors.

\subsection{Modal reduction and frequency sweeps}

Evaluating structural performance across a broad frequency band requires
solving Eq.~\eqref{eq:real_block} repeatedly at many frequency points. For 3D
models with approximately $640\,000$ free degrees of freedom, the cumulative cost of
repetitive full-order factorizations and solves becomes a bottleneck. To
accelerate frequency-sweep post-evaluation, we employ a reduced-order model (ROM)
based on a truncated basis of structural modes; the unreduced finite-element system is referred to as the full-order model (FOM).

A modal basis $\mathbf{\Phi} \in \mathbb{R}^{N_{\mathrm{free}} \times m}$ is
constructed by solving the generalized eigenvalue problem $\mathbf{K} \mathbf{v}
= \lambda \mathbf{M} \mathbf{v}$ for the $m$ lowest modes. We use the
shift-invert Lanczos iteration provided by ARPACK \cite{lehoucq1998arpack}.
To optimize performance, we pre-factorize the shifted stiffness matrix
$(\mathbf{K} - \lambda_{\mathrm{shift}} \mathbf{M})$ using PARDISO's \texttt{factorized}
operator, where $\lambda_{\mathrm{shift}}$ is a small positive shift that improves the numerical
robustness of the shift-invert eigensolution near the lowest elastic modes.
This one-time factorization is then reused throughout the Lanczos iterations to
execute fast linear solves, which keeps the modal-basis construction tractable
for the high-resolution extruded models investigated in this work.

Once the basis is established, the frequency-domain displacement vector
$\mathbf{U}$ is approximated as $\mathbf{U}(\omega) \approx \mathbf{\Phi}
\mathbf{q}(\omega)$. Substituting this into Eq.~\eqref{eq:freq_domain} and
pre-multiplying by $\mathbf{\Phi}^T$ yields a reduced-order system of size
$m \times m$:
\begin{equation}\label{eq:reduced_order}
  \left( \mathbf{\Phi}^T \mathbf{K} \mathbf{\Phi}
  - \omega^2 \mathbf{\Phi}^T \mathbf{M} \mathbf{\Phi}
  + \mathrm{i}\,\omega \mathbf{\Phi}^T \mathbf{C} \mathbf{\Phi}
  + \mathrm{i}\,\eta_s \mathbf{\Phi}^T \mathbf{K} \mathbf{\Phi} \right) \mathbf{q}
  = \mathbf{\Phi}^T \mathbf{F}
\end{equation}
The undamped modal basis diagonalizes $\mathbf{K}$ and $\mathbf{M}$; because the
structural-loss term is proportional to $\mathbf{K}$, it remains diagonal in the
same coordinates. The implementation nonetheless forms and solves the small
$m\times m$ reduced system directly at each frequency. The reported broadband
sweeps use $\mathbf{C}=\mathbf{0}$ and $\eta_s=0.02$. The unified deck
frequency-response sweeps use $m=200$ retained modes.
The reduced displacement is inserted into the
same AIP definition as the full-order model:
\begin{equation}\label{eq:reduced_aip}
  \Pi_{\mathrm{in}}(\omega) \approx
  \frac{1}{2}\,\mathrm{Re}\!\left[
    \mathbf{F}^H \mathbf{\Phi}\dot{\mathbf{q}}(\omega)
  \right]
\end{equation}
A truncated basis under-resolves the static flexibility
$\mathbf{F}^T\mathbf{K}^{-1}\mathbf{F}$. Below the first elastic resonance of the
extruded 3D models ($f_1\approx180$--$560$~Hz) the deck response is
stiffness-controlled, so the active input power is biased
low by the unresolved residual flexibility. We therefore add a static
residual-flexibility (mode-acceleration) correction~\cite{cornwell1983mode}.
The omitted higher modes ($r>m$) are treated as quasi-static over the evaluated
band, leading to the following residual-flexibility correction at each
frequency:
\begin{equation}\label{eq:residual_correction}
  \Delta\Pi(\omega)=\tfrac{1}{2}\,\omega\eta_s\left[
    \mathbf{F}^T\mathbf{K}^{-1}\mathbf{F}
    -\sum_{r\le m}\frac{(\boldsymbol{\phi}_r^T\mathbf{F})^{2}}{k_r}\right],
\end{equation}
where $\mathbf{F}^T\mathbf{K}^{-1}\mathbf{F}$ is the exact static flexibility from
one full-order static solve under the harmonic-force amplitude,
$\boldsymbol{\phi}_r$ is the $r$th retained mode,
$k_r=\boldsymbol{\phi}_r^T\mathbf{K}\boldsymbol{\phi}_r$ is its modal stiffness,
and the sum is the flexibility
already captured by the $m$ retained modes. Adding $\Delta\Pi$ to the truncated
active power removes the low-frequency bias without enlarging the basis.
Once the basis is built, each 3D deck frequency sweep reduces to small dense
complex solves in modal coordinates. Accuracy is checked against selected
full-order PARDISO solves using
\begin{equation}\label{eq:rom_error}
  \epsilon_{\Pi}(\omega) =
  \frac{\left|\Pi_{\mathrm{ROM}}(\omega)-\Pi_{\mathrm{FOM}}(\omega)\right|}
       {\left|\Pi_{\mathrm{FOM}}(\omega)\right|}\times100\% .
\end{equation}
The correction is applied to each deck layout using its residual flexibility.
The 2D frame sweeps are computed by direct re-analysis over 1--500~Hz.

\section{Case Definitions and Numerical Experiments}\label{sec:cases}

The framework is assessed on an engine-supporting panel and a
thruster foundation frame. Both cases use steel as the material with $E=210$~GPa,
$\varrho=7860$~kg/m$^3$, and $\nu=0.3$.

\subsection{Case I: Ship deck panel}\label{sec:case_deck}

The first case represents an engine-supporting panel in a harbour or escort tug.
The dimensions and loading arrangement are informed by the CAT~3516 engine
class, including its published 1.53~m width
\cite{caterpillar2019industrial}. The topology-optimized designs are compared
with stiffened single-skin and X-core double-skin panel references. These reference families are motivated by established stiffened-panel and laser-welded sandwich-panel construction
\cite{putranto2021ultimate, romanoff2007laser}.

\begin{figure}[pos=h]
  \centering
  \includegraphics[width=\columnwidth]{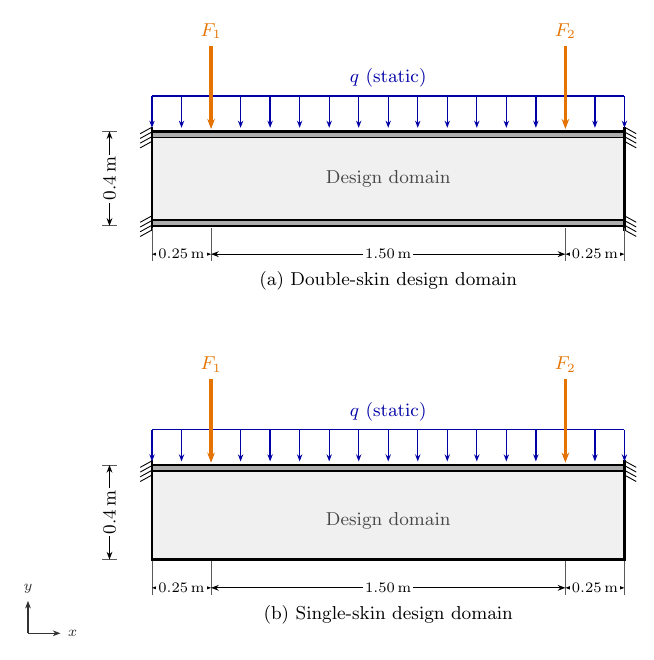}
  \caption{Two-dimensional deck-panel design domains and loading: (a)
    double-skin and (b) single-skin configurations. The gray interiors are
    designable, the black edge strips are passive skins, and the hatched end
    symbols denote clamped supports. The orange arrows indicate the two 15~kN
    harmonic forces, and the blue arrows indicate the uniformly distributed
    12~kN static load.}
  \label{fig:deck_bc_2d}
\end{figure}

Figure~\ref{fig:deck_bc_2d} defines the two-dimensional models. Both use a
$2.0\,\mathrm{m}\times0.4\,\mathrm{m}$ domain, a $320\times64$ plane-stress
mesh with thickness $t=0.01$~m. The
double-skin designs have four passive element rows forced to be solid at both the top and bottom,
with the ends of both skins clamped. The single-skin designs have four passive-solid element rows at the top, whose ends are clamped. The remaining elements are designable, and
the total solid fraction, including passive material, is $v_f=0.30$.

The deck is optimized at 100~Hz. A total 30~kN harmonic load is split
equally between two top-surface locations, $x=0.25$~m and $x=1.75$~m, labelled
$F_1$ and $F_2$ in Figure~\ref{fig:deck_bc_2d}.
Their 1.5~m separation follows the width of the engine supports. A separate total static load of 12~kN, labelled
$q$ in Figure~\ref{fig:deck_bc_2d}, is distributed uniformly downward
along the top boundary and defines the non-harmonic service load used for the
static-compliance evaluation. The same prescribed amplitudes are used for all deck variants, so their responses
are compared under a common load case. Off-design response
is evaluated by frequency sweeps.

Four TO variants are considered:
\emph{Unrestricted} (cone density filter only), \emph{Manufacturing-aware}
(Helmholtz PDE filter with Heaviside three-field projection),
\emph{Stress-aware} (cone filter with the $p$-norm stress term), and
\emph{Stress~+~Manufacturing} (Helmholtz filter, Heaviside projection, and
stress term). All use KATO with
$w_{\mathrm{AIP}}=0.9$. The standard runs use 120 optimization steps; the
stress-aware runs use 180 steps.

\begin{figure}[pos=h]
  \centering
  \includegraphics[width=\columnwidth]{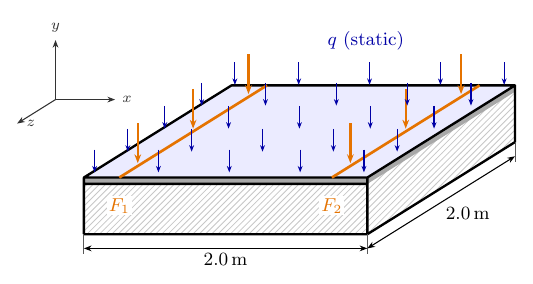}
  \caption{Three-dimensional re-analysis model formed by extruding the
    optimized deck cross sections by 2.0~m. The two 15~kN harmonic loads
    $F_1$ and $F_2$ are distributed along the extrusion direction at
    $x=0.25$~m and $x=1.75$~m, respectively. The 12~kN static load $q$ is
    distributed over the full top surface, and the hatched lateral faces
    denote clamped boundaries.}
  \label{fig:deck_bc_3d}
\end{figure}

Once the 2D designs are obtained, each binary cross-section is extruded
$2.0$~m in the $z$ direction
(Figure~\ref{fig:deck_bc_3d}). Material nodes on the four lateral boundary
faces are clamped, while the cross-section faces remain free. The 3D
re-analysis retains the nominal 30~kN harmonic and 12~kN static resultants
used in 2D: the two 15~kN harmonic components are distributed along the
extrusion direction at $x=0.25$~m and $x=1.75$~m, and the static resultant is
distributed over the full top surface. Because the 2D cross-section and
finite-depth 3D models distribute these resultants differently, their metrics
are compared within each model.

\begin{figure}[pos=h]
  \centering
  \includegraphics[width=\columnwidth]{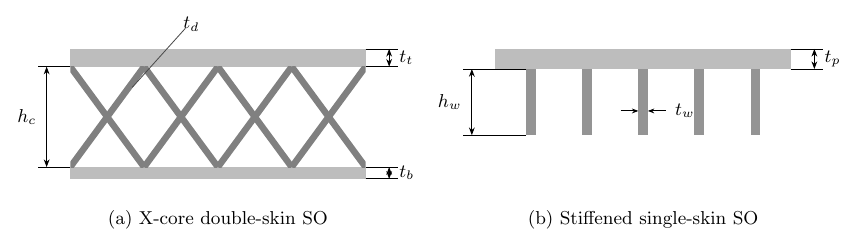}
  \caption{Cross-sections of the size-optimized engineering references:
    (a) X-core double-skin and (b) stiffened single-skin panels.}
  \label{fig:so_models}
\end{figure}

The engineering references (stiffened single-skin and X-core double-skin panels) retain their prescribed cross-sectional topology while their member
sizes and spacings are optimized (Figure~\ref{fig:so_models}). The X-core double-skin reference varies
the face plate thicknesses, diagonal webs, core height, and number of units in the core. The stiffened
single-skin reference varies the plate thickness, web thickness and height, and number of webs. These structures are loaded and constrained using the same approach as for the TO variants shown in Figure~\ref{fig:deck_bc_2d}. An
exhaustive grid search over the ranges in Table~\ref{tab:so_bounds} screens
candidates against the approximate volume-matching band
$0.285\leq v_f\leq0.315$ and retains the lowest-AIP design in that band. The X-core reference is compared with the double-skin TO family, whereas
the stiffened reference is compared with the single-skin TO family.
Both references are represented as binary geometries in the 2D analysis and
are subsequently extruded using the same 3D re-analysis procedure as the
corresponding TO family. Together with the four variants in each TO family,
the case study compares ten designs in total.

\begin{table}[pos=h]
  \centering
  \caption{Parameter ranges for the size-optimized cross-sections of the
    X-core double-skin and stiffened single-skin panels.}
  \label{tab:so_bounds}
  \footnotesize
  \begin{tabular}{@{}llccc@{}}
    \toprule
    & Parameter & \makecell{Range\\(cells)} & \makecell{Range\\(mm)} & Selected \\
    \midrule
    \multirow{5}{*}{\rotatebox[origin=c]{90}{X-core}}
      & Top-face thickness $t_t$     & 2--6   & 12.5--37.5  & $37.5$~mm \\
      & Bottom-face thickness $t_b$  & 2--6   & 12.5--37.5  & $18.75$~mm \\
      & Diagonal-web thickness $t_d$ & 2--6   & 12.5--37.5  & $18.75$~mm \\
      & Core height $h_c$            & 16--60 & 100--375    & $343.75$~mm \\
      & Number of X-units $n_u$      & 3--16  & --          & $8$ \\
    \midrule
    \multirow{4}{*}{\rotatebox[origin=c]{90}{Stiffened}}
      & Top-plate thickness $t_p$ & 2--8   & 12.5--50.0  & $50.0$~mm \\
      & Web thickness $t_w$       & 2--5   & 12.5--31.25 & $31.25$~mm \\
      & Web height $h_w$          & 20--60 & 125--375    & $200$~mm \\
      & Number of webs $n_w$      & 8--24  & --          & $24$ \\
    \bottomrule
  \end{tabular}
\end{table}

\subsection{Case II: Thruster foundation frame}\label{sec:case_frame}

\begin{figure}[pos=h]
  \centering
  \includegraphics[width=\columnwidth]{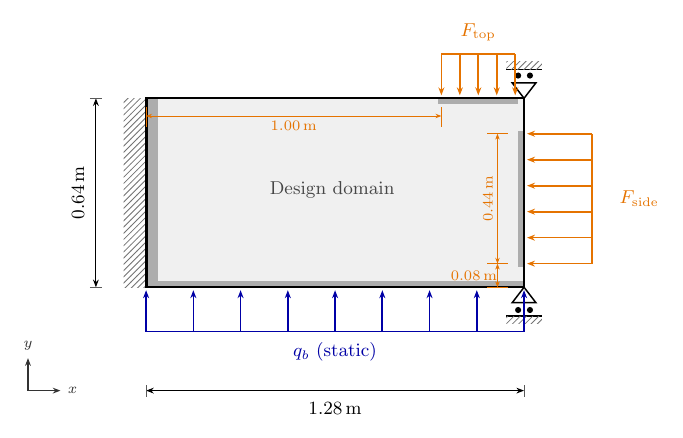}
  \caption{Design domain, boundary conditions, passive-solid regions, and
    loading for the thruster foundation frame. The gray regions denote
    passive-solid connection and load-introduction zones. The orange arrows
    mark the 30~kN side and top load components, applied over
    $y=0.08$--$0.52$~m and $x=1.00$--$1.25$~m, respectively. The blue arrows
    indicate the uniformly distributed 12~kN static uplift $q_{\mathrm{b}}$
    over the full bottom edge.}
  \label{fig:frame_bc}
\end{figure}

The second case represents a side frame in an azimuthing-thruster foundation. Its engineering context is the \emph{Seaspan Raven}, a Robert Allan
Ltd RAstar~2800 escort tug with twin CAT~3516B engines driving Rolls--Royce
US~205~CP azimuthing Z-drives with 2.4~m propellers
\cite{robertallan2011seaspan}. Figure~\ref{fig:frame_bc} shows the frame design domain, represented by a $128\times64$ plane-stress mesh ($1.28\,\mathrm{m}\times0.64\,\mathrm{m}$) with out-of-plane thickness $t=0.01$~m. The left-hand side is fully restrained, and two short right-hand side
pads provide local vertical ($u_y=0$) support. The $18$~Hz harmonic excitation
combines an inward side-thruster load $F_{\mathrm{side}}$ distributed over the right-hand boundary
($y=0.08$--$0.52$~m) with a harmonic load $F_{\mathrm{top}}$ distributed over
$x=1.00$--$1.25$~m on the top boundary. The two
harmonic components are represented by a real force vector and applied in
phase. The static-compliance load case combines the corresponding side and top
load components with a uniform static uplift $q_{\mathrm{b}}$ on the bottom edge. The design
frequency represents the blade-passing frequency
$f_{\mathrm{BPF}}=N_b n/60$ of a four-bladed propeller ($N_b$ = 4) operating at
approximately $n$ = 270~rpm.
The harmonic loads are each 30~kN, representing horizontal thrust
fluctuation and local vertical machinery excitation, respectively. The
corresponding static side and top components retain these amplitudes, and the
12~kN bottom static uplift represents an equivalent hydrostatic action.
The gray regions in Figure~\ref{fig:frame_bc} mark the passive-solid left
connection column, the side and top load-introduction zones, and the bottom
load rows. These elements are held at full density; the design volume fraction
is adjusted so that the total volume fraction is $v_f=0.45$.

In this case, GCMMA \cite{svanberg1987method, svanberg2002class} is used as a density-based benchmark for KATO. Four variant labels are used for
each method:
\emph{Unrestricted}, \emph{Manufacturing-aware}, \emph{Stress-aware} and
\emph{Stress~+~Manufacturing}.
All runs use $w_{\mathrm{AIP}}=0.8$, 160 optimization steps, PARDISO with six
threads, and the binary protocol below. The Unrestricted and Stress-aware
GCMMA variants use a cone filter with $r_{\min}=3$ elements. Variants with
manufacturing control use the same Helmholtz filter in both methods. KATO then
applies the single-field Heaviside projection described in
Section~\ref{sec:density_pipeline}, whereas GCMMA handles the volume fraction
through its constrained optimization update.

An additional \emph{Unrestricted} optimization is performed at 300~Hz, close to the
306~Hz response peak of the uniform starting field. The same control parameters are used.

\subsection{Binary post-evaluation protocol}

After each TO convergence, the continuous density field is binarized by a
volume-preserving threshold: each element with
$\rho_i \ge \rho^{*}$ is set solid ($1$) and the remainder void ($0$). The cut
level $\rho^{*}$ is selected from the final density values to give the closest
element-wise realization of $v_f$. In the 2D re-analysis, void elements are assigned
the numerical ersatz density $\rho_{\mathrm{ersatz}}=10^{-3}$ to maintain matrix
conditioning. The extruded 3D models retain the binary geometry and use the
minimum stiffness and mass densities in Eqs.~\eqref{eq:simp} and
\eqref{eq:mass_interp}. AIP, compliance, stress, and connectivity are recomputed
on the binary models. The continuous-field diagnostic $M_{nd}$ is evaluated over
the full frame domain, including the common passive regions.
Connectivity is quantified by $N_c$, the number of four-neighbor connected
solid components in the volume-matched binary field; $N_c=1$ denotes a single
connected topology.

\section{Results}\label{sec:results}

\subsection{Ship deck panel}

Table~\ref{tab:deck_comparison} compares the optimized panel cross-sections at the 100~Hz design frequency. All topology-optimized variants reduce AIP relative to their corresponding engineering references. The double-skin designs yield 130.02--131.22~dB, compared with 136.30~dB for the X-core double-skin reference. The single-skin designs yield 135.03--135.99~dB, compared with 147.99~dB for the stiffened single-skin reference. The largest separation occurs in the single-skin family: the Unrestricted and Manufacturing-aware layouts give static compliances of 11.4 and 11.8~N\,m, compared with 542~N\,m for the stiffened single-skin reference. In the double-skin family, the corresponding values are 6.12 and 6.65~N\,m, compared with 11.3~N\,m for the X-core reference.
The Stress-aware variants trade some stiffness for lower dynamic stress. The
double-skin designs attain the lowest AIP because their top and bottom skins
provide a closed load path with high bending efficiency under through-thickness
vibration, whereas the single-skin designs rely on the top plate.

\begin{table*}[pos=h]
  \centering
  \caption{Performance at 100~Hz for the eight KATO panel designs and two
    size-optimized engineering references. Stress values use the combined dynamic von Mises
    measures, and the solid fraction includes passive material. Bold values
    identify the best KATO result within each structural family.}
  \label{tab:deck_comparison}
  \footnotesize
  \setlength{\tabcolsep}{5pt}
  \begin{tabular}{@{}lccccc@{}}
    \toprule
    Structure & \makecell{AIP@100\\(dB)} & \makecell{Static\\comp. (N\,m)} &
    \makecell{Dyn.\\$p$-norm (MPa)} & \makecell{Dyn.\\max (MPa)} &
    \makecell{Solid\\frac.} \\
    \midrule
    \text{Double-skin designs:} \\
    Unrestricted                    & 131.14 & \textbf{6.12} & 41.8 & 97.6 & 0.300 \\
    Manufacturing-aware             & 131.22 & 6.65 & 35.4 & 89.1 & 0.300 \\
    Stress-aware                    & \textbf{130.02} & 46.5 & \textbf{29.2} & \textbf{67.8} & 0.300 \\
    Stress + Manufacturing          & 130.09 & 22.8 & 29.7 & 73.0 & 0.300 \\
    \textit{X-core double-skin (SO)} & \textit{136.30} & \textit{11.3} & \textit{69.0} & \textit{183.2} & \textit{0.291} \\
    \midrule
    \text{Single-skin designs:} \\
    Unrestricted                         & 135.28 & \textbf{11.4} & 54.7 & 139.5 & 0.300 \\
    Manufacturing-aware                  & 135.20 & 11.8 & 54.7 & 139.8 & 0.300 \\
    Stress-aware                         & 135.99 & 42.0 & \textbf{49.1} & \textbf{118.6} & 0.300 \\
    Stress + Manufacturing               & \textbf{135.03} & 64.4 & 49.6 & 125.5 & 0.300 \\
    \textit{Stiffened single-skin (SO)} & \textit{147.99} & \textit{542.0} & \textit{317.1} & \textit{841.5} & \textit{0.312} \\
    \bottomrule
  \end{tabular}
\end{table*}

\begin{figure*}[pos=h]
  \centering
  \includegraphics[width=\textwidth]{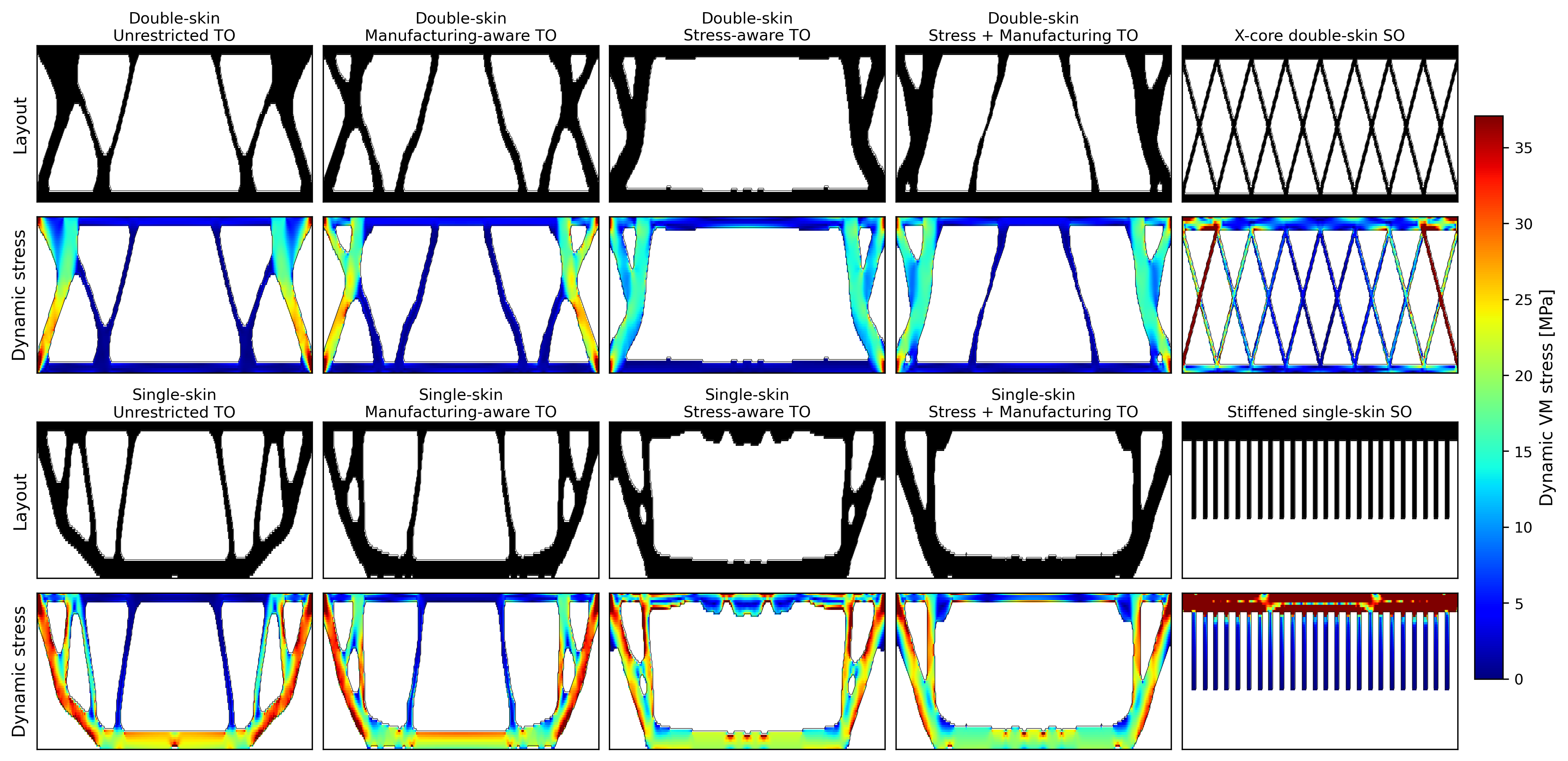}
  \caption{KATO-optimized cross-sections and corresponding 100~Hz dynamic
    von Mises stress fields for the double-skin and single-skin designs. Each
    family is grouped with its corresponding size-optimized reference. The
    common stress scale is capped at the 98th percentile of the eight KATO
    layouts; values above this limit are saturated.}
  \label{fig:deck_2d_structures}
\end{figure*}

For the double-skin family, the Stress-aware variant reduces the dynamic von
Mises stress $p$-norm from 41.8~MPa (Unrestricted) to 29.2~MPa. In the
single-skin family, it reduces the same measure from 54.7~MPa to 49.1~MPa.
These stress reductions come
with higher static compliance, while the AIP changes by less than 1~dB in both
families. The Manufacturing-aware variant also lowers the dynamic stress in the
double-skin family (from 41.8 to 35.4~MPa) by suppressing thin members through
feature-size control. In the single-skin family, the Manufacturing-aware
variant converges to nearly the same layout as the Unrestricted variant,
indicating that the unrestricted topology is already compatible with the
imposed feature-size control. Their AIP and stress $p$-norm values are
consequently nearly identical (135.2 versus 135.3~dB and 54.7~MPa in both). The
two size-optimized references have slightly different solid fractions (0.291 and
0.312) because their parametric search returns discrete sizing; the stiffened
single-skin reference channels the load through a few webs and reaches a much higher
stress $p$-norm (317~MPa) and AIP. Figure~\ref{fig:deck_2d_structures} compares the binary layouts and dynamic von
Mises stress fields. 

\begin{figure*}[pos=p]
  \centering
  \includegraphics[height=0.90\textheight,keepaspectratio]{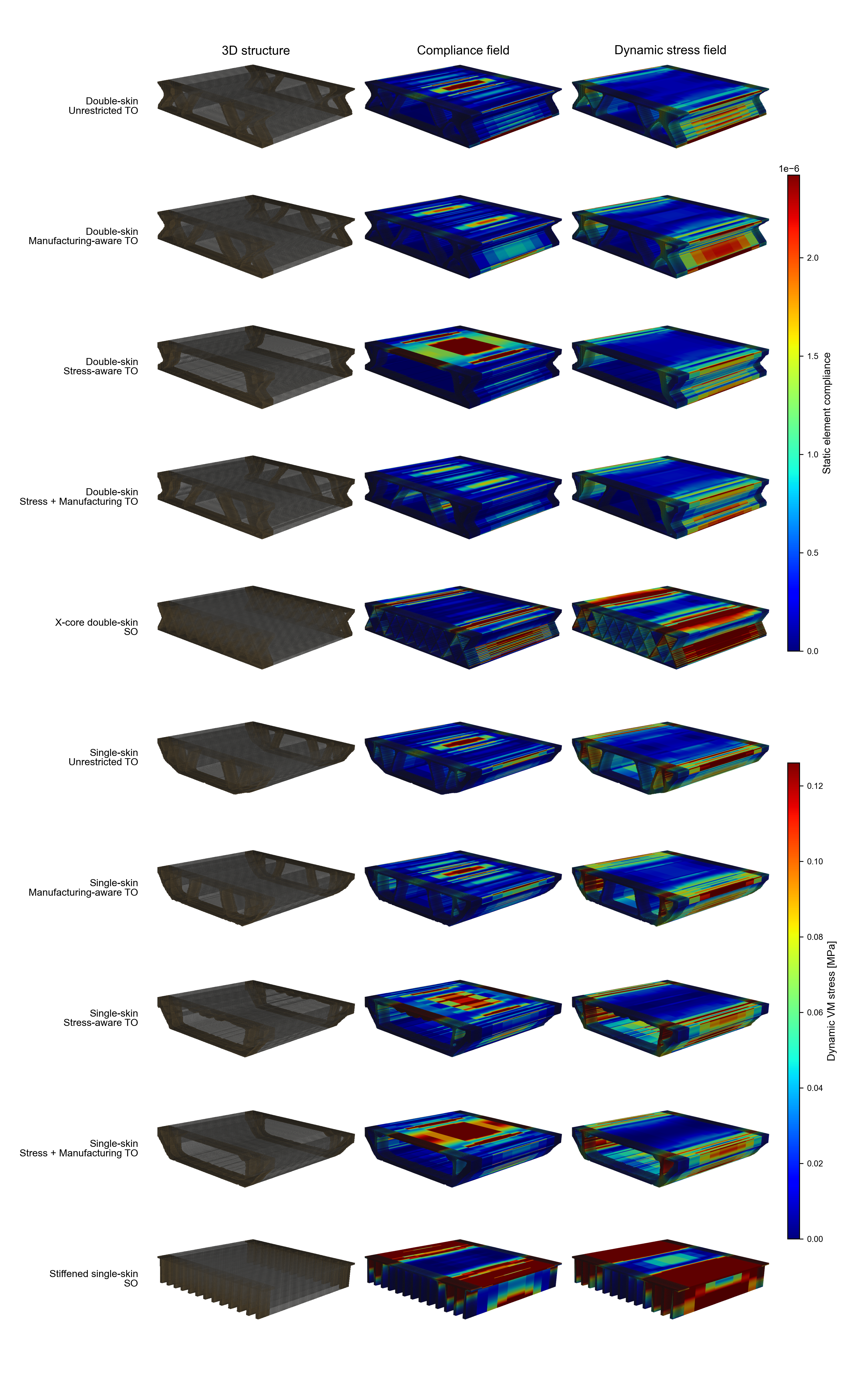}
  \caption{Extruded three-dimensional models of the eight KATO designs and two
    size-optimized references. Columns show the optimized structure,
    element-wise static-compliance contribution, and 100~Hz combined dynamic von Mises
    stress field. The compliance and stress scales are capped at the 98th
    percentile of the eight KATO layouts; values above these limits are
    saturated.}
  \label{fig:deck_3d_structures}
\end{figure*}

Figure~\ref{fig:deck_3d_structures} shows the extruded 3D
models with static-compliance and dynamic-stress fields, and the KATO layouts
remain connected and retain their load paths after extrusion. The
stiffened single-skin reference concentrates stress and strain energy (compliance) in a small number
of webs. Because the 2D and 3D models distribute the load through different
thicknesses, their absolute metrics are compared within each model.

\begin{figure}[pos=h]
  \centering
  \includegraphics[width=\columnwidth]{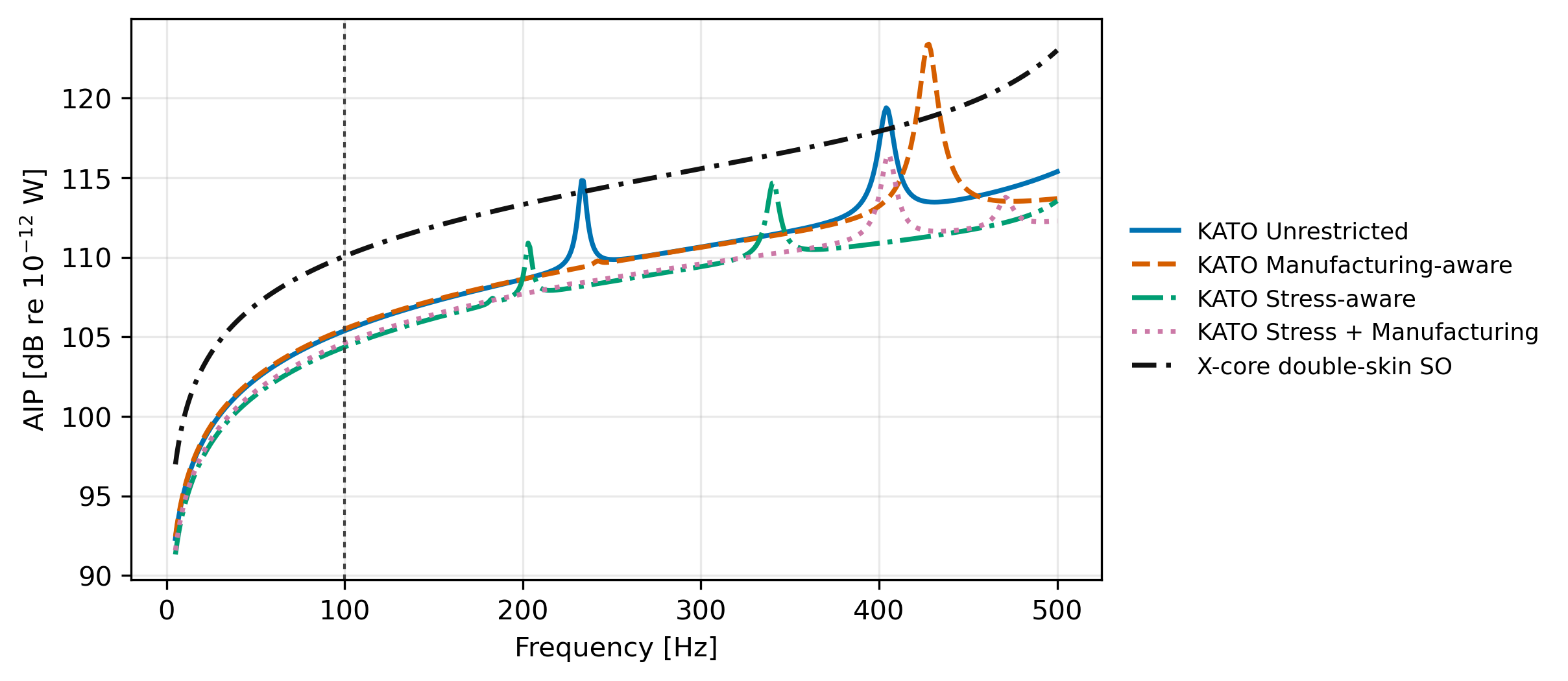}
  \caption{AIP frequency responses for extruded models over 5--500~Hz for the
    four double-skin KATO designs and the size-optimized X-core double-skin
    reference. The vertical dashed line marks the 100~Hz design frequency.}
  \label{fig:deck_frf_sandwich}
\end{figure}

\begin{figure}[pos=h]
  \centering
  \includegraphics[width=\columnwidth]{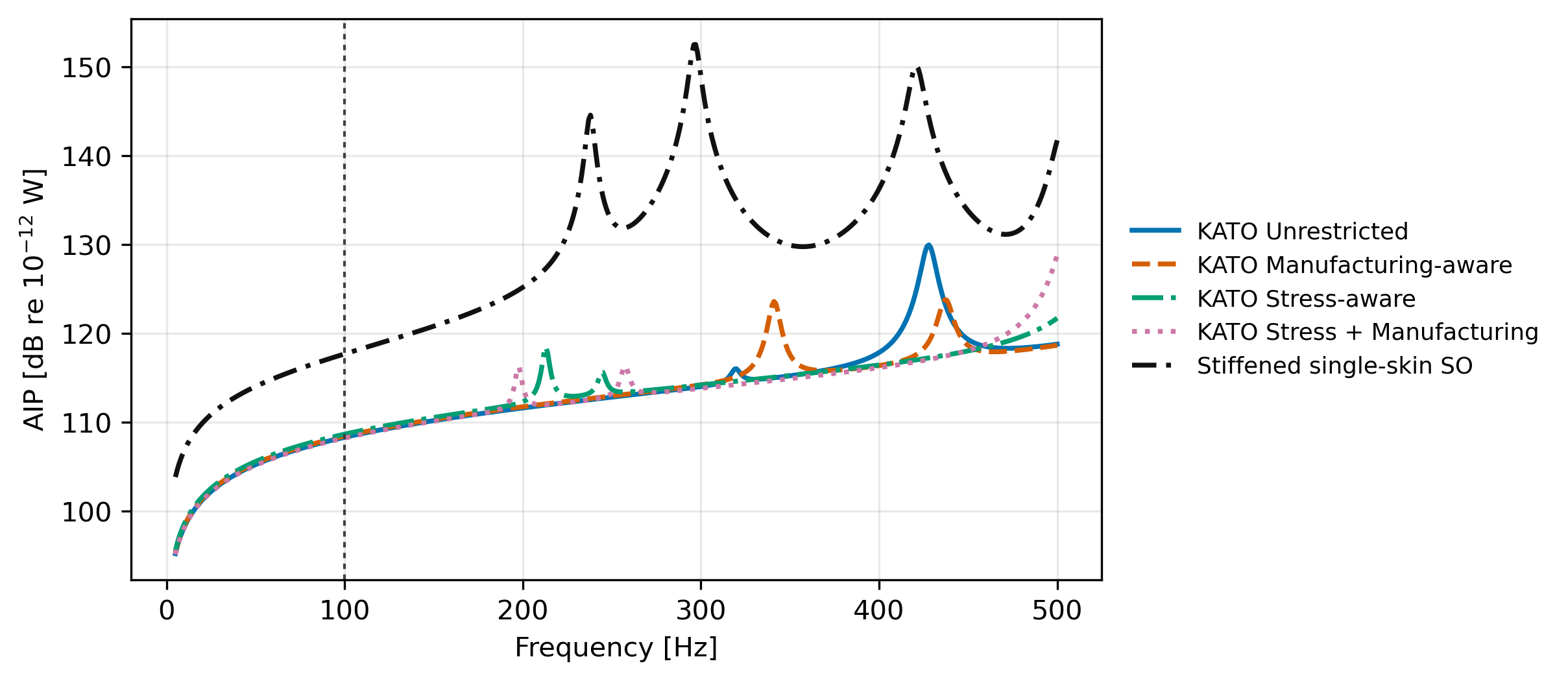}
  \caption{AIP frequency responses for extruded models over 5--500~Hz for the
    four single-skin KATO designs and the size-optimized stiffened single-skin
    reference. The vertical dashed line marks the 100~Hz design frequency.}
  \label{fig:deck_frf_deck}
\end{figure}

Figure~\ref{fig:deck_frf_sandwich} shows that all four double-skin KATO designs
retain their design-frequency advantage after 3D extrusion, with AIP values
4.6--5.7~dB below the X-core reference at 100~Hz. Their first modal frequencies
and maximum-response frequencies vary with topology, producing distinct
off-design responses while remaining below the reference over most of the
sweep.

Figure~\ref{fig:deck_frf_deck} shows a larger design-frequency separation in the case of the single-skin designs: the
four KATO layouts are 9.0--9.5~dB below the stiffened reference at
100~Hz. The reference also exhibits higher resonant peaks over the sweep,
whereas the KATO layouts maintain lower responses with topology-dependent peak
locations. Both frequency responses are computed with the corrected 200-mode
model.

Table~\ref{tab:deck_3d_validation} gives the corresponding quantitative
3D metrics. Here $C_s$ is static compliance,
$\sigma_p$ is the dynamic stress $p$-norm, $f_1$ is the first elastic modal
frequency, and $f_{\max}$ is the frequency of maximum AIP within the
5--500~Hz sweep. At 100~Hz the topology-optimized layouts
remain below both engineering baselines in AIP and compliance. Their resonant
responses are layout dependent across the sweep.

\begin{table}[pos=h]
  \centering
  \caption{Performance of the eight extruded KATO deck models and two
    size-optimized references. AIP and the dynamic stress $p$-norm are evaluated
    at 100~Hz; $C_s$ is static compliance, $f_1$ is the first elastic modal
    frequency, and $f_{\max}$ is the frequency of maximum AIP within the
    5--500~Hz sweep.}
  \label{tab:deck_3d_validation}
  \footnotesize
  \setlength{\tabcolsep}{5pt}
  \begin{tabular}{@{}lrrrrr@{}}
    \toprule
    Structure & AIP (dB) & $C_s$ (N\,m) & $\sigma_p$ (MPa) & $f_1$ (Hz) & $f_{\max}$ (Hz) \\
    \midrule
    \text{Double-skin designs:} \\
    Unrestricted                    & 105.38 & $5.43\times10^{-3}$ & 0.148 & 233 & 404 \\
    Manufacturing-aware             & 105.49 & $5.58\times10^{-3}$ & 0.091 & 242 & 428 \\
    Stress-aware                    & 104.36 & $4.32\times10^{-3}$ & 0.069 & 183 & 341 \\
    Stress + Manufacturing          & 104.61 & $4.58\times10^{-3}$ & 0.070 & 226 & 405 \\
    X-core double-skin (SO)         & 110.06 & $1.59\times10^{-2}$ & 0.197 & 563 & 500$^\dagger$ \\
    \midrule
    \text{Single-skin designs:} \\
    Unrestricted                    & 108.28 & $1.05\times10^{-2}$ & 0.107 & 319 & 428 \\
    Manufacturing-aware             & 108.42 & $1.08\times10^{-2}$ & 0.107 & 341 & 437 \\
    Stress-aware                    & 108.64 & $1.14\times10^{-2}$ & 0.109 & 213 & 500$^\dagger$ \\
    Stress + Manufacturing          & 108.20 & $1.03\times10^{-2}$ & 0.095 & 198 & 500$^\dagger$ \\
    Stiffened single-skin (SO)      & 117.67 & $7.64\times10^{-2}$ & 0.655 & 238 & 297 \\
    \bottomrule
  \end{tabular}
  \par\smallskip
  \parbox{0.96\textwidth}{\footnotesize
    $^\dagger$ Maximum at the upper sweep boundary.}
\end{table}

\subsection{Thruster foundation frame: KATO versus GCMMA}

\begin{figure*}[pos=h]
  \centering
  \includegraphics[width=\textwidth]{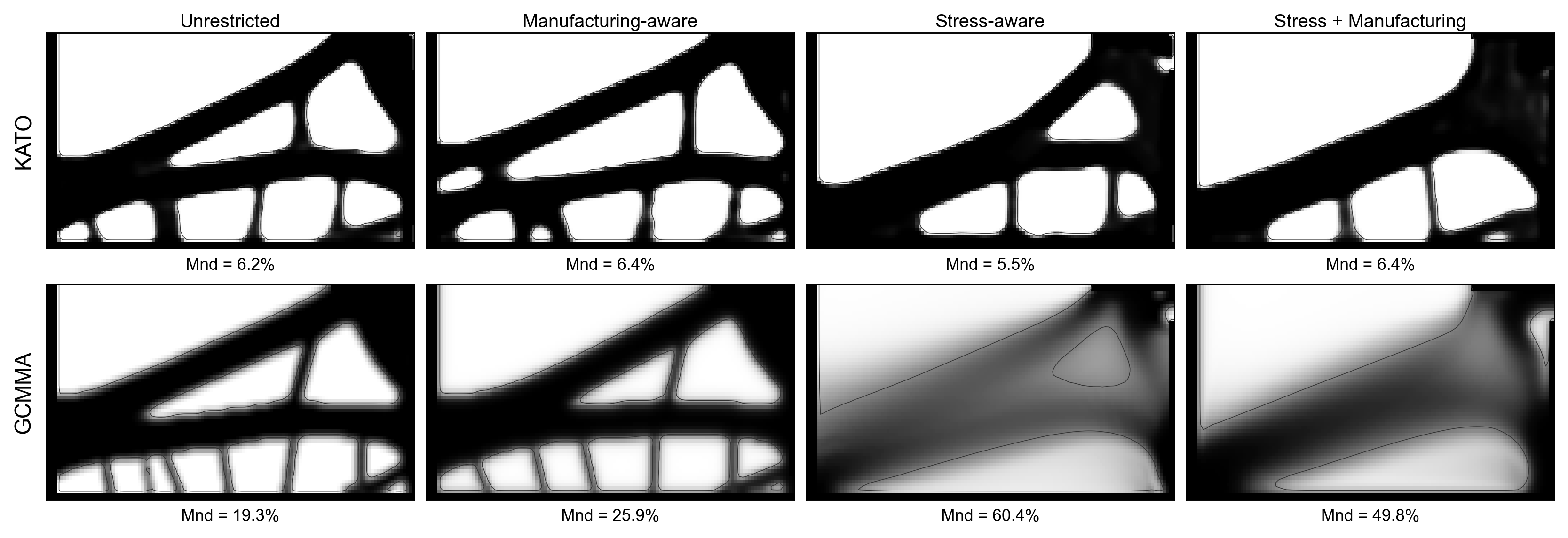}
  \caption{Continuous final density fields before volume-preserving
    thresholding for the four KATO variants (top) and four GCMMA variants
    (bottom). The reported $M_{nd}$ values quantify pre-threshold
    non-discreteness.}
  \label{fig:frame_grayfield}
\end{figure*}

Table~\ref{tab:frame_comparison} shows the performance of the foundation frame for different TO variants and compares KATO with GCMMA. The associated computational cost is shown as well. The $M_{nd}$ values characterize the continuous final
fields. Across all four variants, KATO achieves 126.2--126.9~dB AIP at a binary static compliance of 3.6--3.8~N\,m and a dynamic stress $p$-norm of
27.5--38.2~MPa. The GCMMA layouts reach comparable AIP for the Unrestricted,
Manufacturing-aware, and Stress-aware variants (125.8--126.7~dB) but at a binary
static compliance that is 22--36$\times$ larger (80--135~N\,m). Both
continuous stress-constrained GCMMA solutions satisfy the 26~MPa limit, with
static compliances of 2.59 and 2.68~N\,m. After thresholding, the Stress-aware
and Stress~+~Manufacturing designs reach stress $p$-norms of 33.20 and
218.77~MPa and static compliances of 134.55 and 71.41~N\,m, respectively. This
continuous-to-binary change, together with the higher non-discreteness of the
GCMMA fields ($M_{nd}=19$--$60\%$; Table~\ref{tab:frame_comparison} and
Figure~\ref{fig:frame_grayfield}), shows how thresholding removes gray load
paths and degrades the final stiffness. The KATO fields are closer to binary
($M_{nd}=5.5$--$6.4\%$), and all four variants retain one connected component
and low compliance after thresholding. The stress-aware KATO variants also
reduce the dynamic stress $p$-norm from 38.2 to 27.5~MPa.

\begin{table*}[pos=h]
  \centering
  \caption{Performance and optimization cost for the four KATO and four GCMMA
    frame variants at 18~Hz. The non-discreteness measure $M_{nd}$ is evaluated
    before thresholding; AIP, compliance, stress, and the connected-component
    count $N_c$ use the post-processed designs. Timing and peak resident set
    size use six-thread cached PARDISO.}
  \label{tab:frame_comparison}
  \footnotesize
  \setlength{\tabcolsep}{3pt}
  \begin{tabular}{@{}llccccccccc@{}}
    \toprule
    Method & Variant & \makecell{AIP\\(dB)} & \makecell{Static\\comp. (N\,m)} &
    \makecell{Dyn.\\$p$-norm (MPa)} & \makecell{Max\\(MPa)} & \makecell{$M_{nd}$\\(\%)} & $N_c$ &
    \makecell{Run\\(s)} & \makecell{Eff.\\evals} & \makecell{Peak\\RSS (MB)} \\
    \midrule
    \multirow{4}{*}{KATO}
          & Unrestricted           & 126.19 & \textbf{3.60} & 38.22 & 105.0 & 6.2 & 1 & \textbf{49.2} & 160 & 1651 \\
          & Manufacturing-aware    & 126.24 & 3.62          & 36.05 & 99.3  & 6.4 & 1 & 50.7          & 160 & 1660 \\
          & Stress-aware           & 126.93 & 3.78          & 29.49 & 78.0  & \textbf{5.5} & 1 & 56.0          & 160 & 1739 \\
          & Stress + Manufacturing & 126.50 & 3.72          & \textbf{27.53} & \textbf{74.9} & 6.4 & 1 & 57.2 & 160 & 1730 \\
    \midrule
    \multirow{4}{*}{GCMMA}
          & Unrestricted           & 125.84 & 95.62  & 36.45  & 100.4 & 19.3 & 3 & 52.6 & 382  & \textbf{598} \\
          & Manufacturing-aware    & \textbf{125.83} & 80.38 & 36.32 & 100.1 & 25.9 & 2 & \textbf{51.6} & 362  & 612 \\
          & Stress-aware           & 126.68 & 134.55 & 33.20  & 83.2  & 60.4 & 1 & 581.1 & 1227 & 604 \\
          & Stress + Manufacturing & 134.87 & 71.41  & 218.77 & 525.1 & 49.8 & 1 & 366.1 & 1112 & 614 \\
    \bottomrule
  \end{tabular}
\end{table*}

Table~\ref{tab:frame_comparison} also reports the optimization cost after the
shared sparse-solver backend acceleration. With 160 outer iterations on a
six-thread cached PARDISO backend, the KATO runs complete in 49--57~s. The GCMMA
Unrestricted and Manufacturing-aware runs have comparable wall time (52--53~s),
whereas the stress-constrained GCMMA runs require 366--581~s because each outer
step solves several inner convex subproblems with conservative GCMMA
approximations. The number of effective evaluations, i.e., total finite-element
objective and sensitivity evaluations, therefore rises to 1112--1227 in the
stress-constrained GCMMA runs, against the fixed 160 for KATO. The neural
reparameterized KATO formulation gives its clearest runtime advantage in these
constrained variants, requiring 160 effective evaluations compared with
1112--1227 for GCMMA, at the cost of a higher peak memory footprint (about
1.65--1.74~GB versus 0.60--0.61~GB), reflecting the autograd graph and generator
network held in memory.

\begin{figure}[pos=h]
  \centering
  \includegraphics[width=\columnwidth]{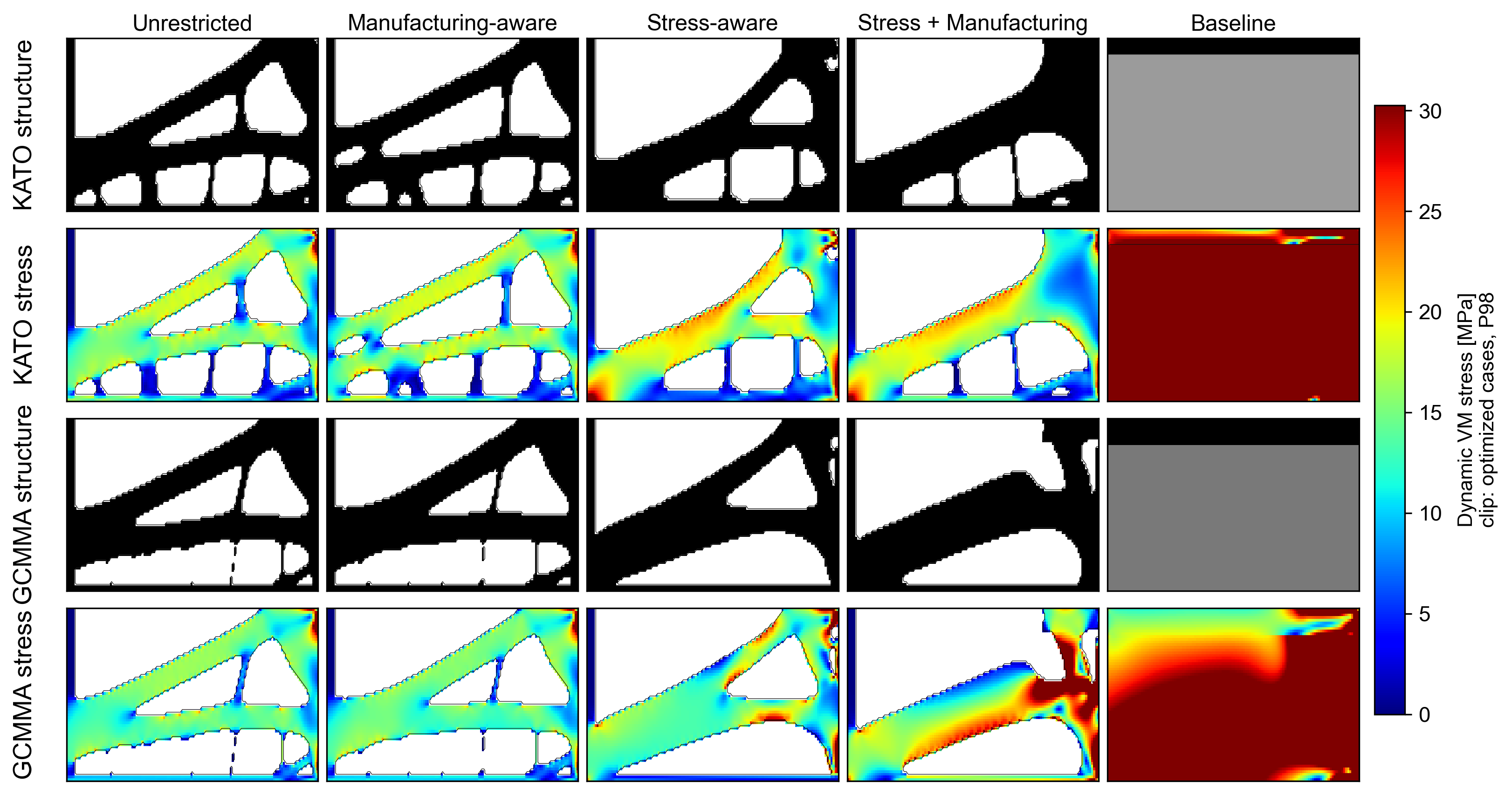}
  \caption{Optimized designs and 18~Hz combined dynamic von Mises stress fields for the
    four KATO variants and the corresponding GCMMA variants. The upper and lower
    pairs of rows show KATO and GCMMA, respectively; the rightmost column shows
    the uniform baseline with a top flange. The stress scale is clipped at the 98th percentile of
    the optimized cases.}
  \label{fig:frame_field_grid}
\end{figure}

Figure~\ref{fig:frame_field_grid} shows that all four KATO variants retain
continuous diagonal and horizontal load paths after thresholding, consistent
with their single connected component. Stress regularization redistributes
stress through broader primary members and lowers the KATO stress $p$-norm,
whereas several GCMMA layouts contain disconnected members or pronounced stress
concentrations.

Figure~\ref{fig:frame_grayfield} shows sharper material boundaries and
consistently low non-discreteness for KATO ($M_{nd}=5.5$--$6.4\%$), compared
with $19.3$--$60.4\%$ for GCMMA. The extensive gray regions in the
stress-constrained GCMMA fields explain their greater sensitivity to
thresholding and the associated changes in binary compliance and stress.

\begin{figure}[pos=h]
  \centering
  \includegraphics[width=\columnwidth]{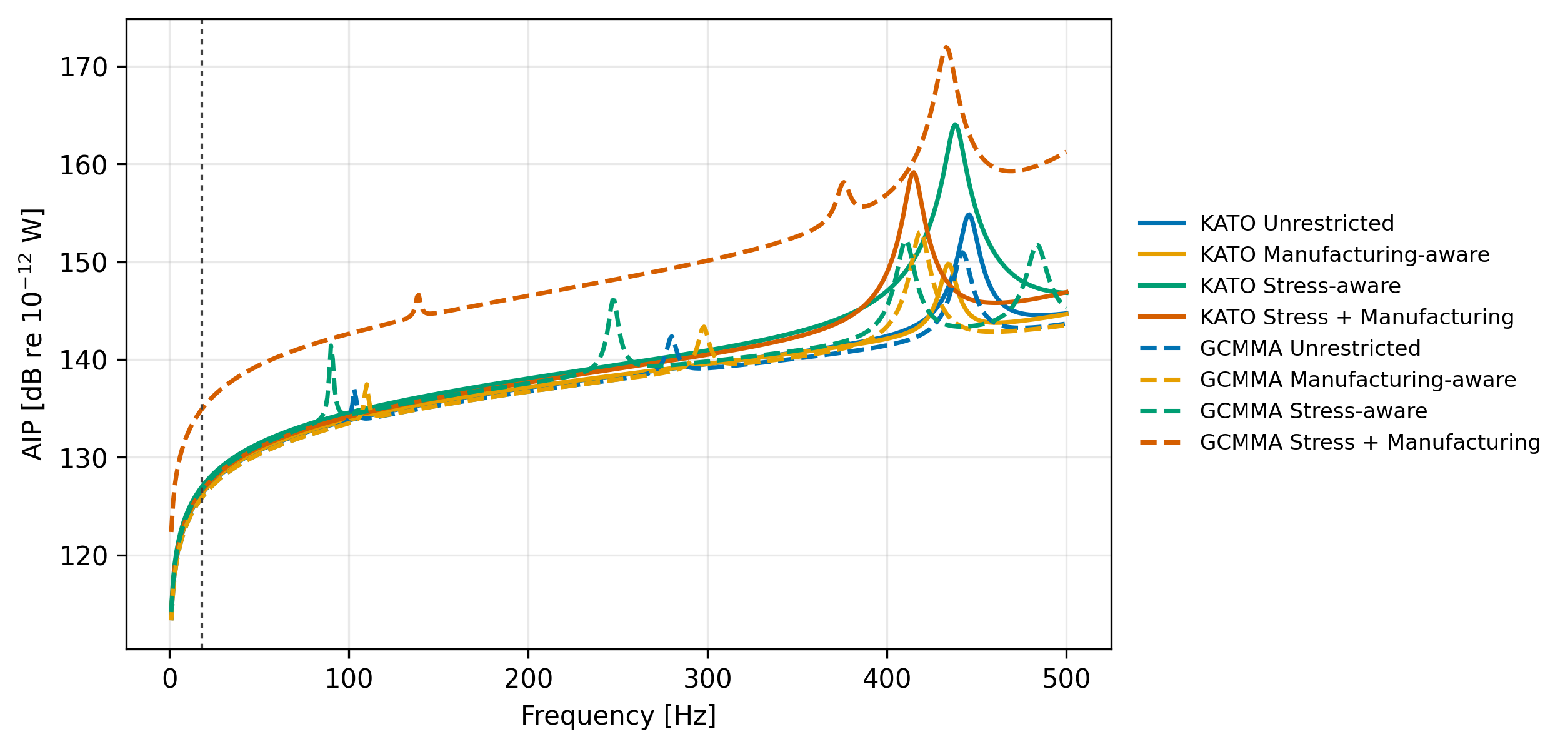}
  \caption{AIP frequency responses over 1--500~Hz for the optimized 18~Hz frame
    designs obtained with the four KATO and four GCMMA variants. The vertical
    dashed line marks the design frequency.}
  \label{fig:frame_frf}
\end{figure}

Figure~\ref{fig:frame_frf} confirms that KATO and the three comparable GCMMA
variants reach closely matched AIP at the 18~Hz design frequency, while the
GCMMA Stress~+~Manufacturing result remains higher. Above the design frequency,
the responses separate and develop load-participating peaks between 410 and
446~Hz, demonstrating the effect of the final topology on off-design dynamic
performance.

\subsection{Near-resonant frame test at 300 Hz}\label{sec:frame_300hz}

\begin{figure}[pos=h]
  \centering
  \includegraphics[width=0.92\textwidth]{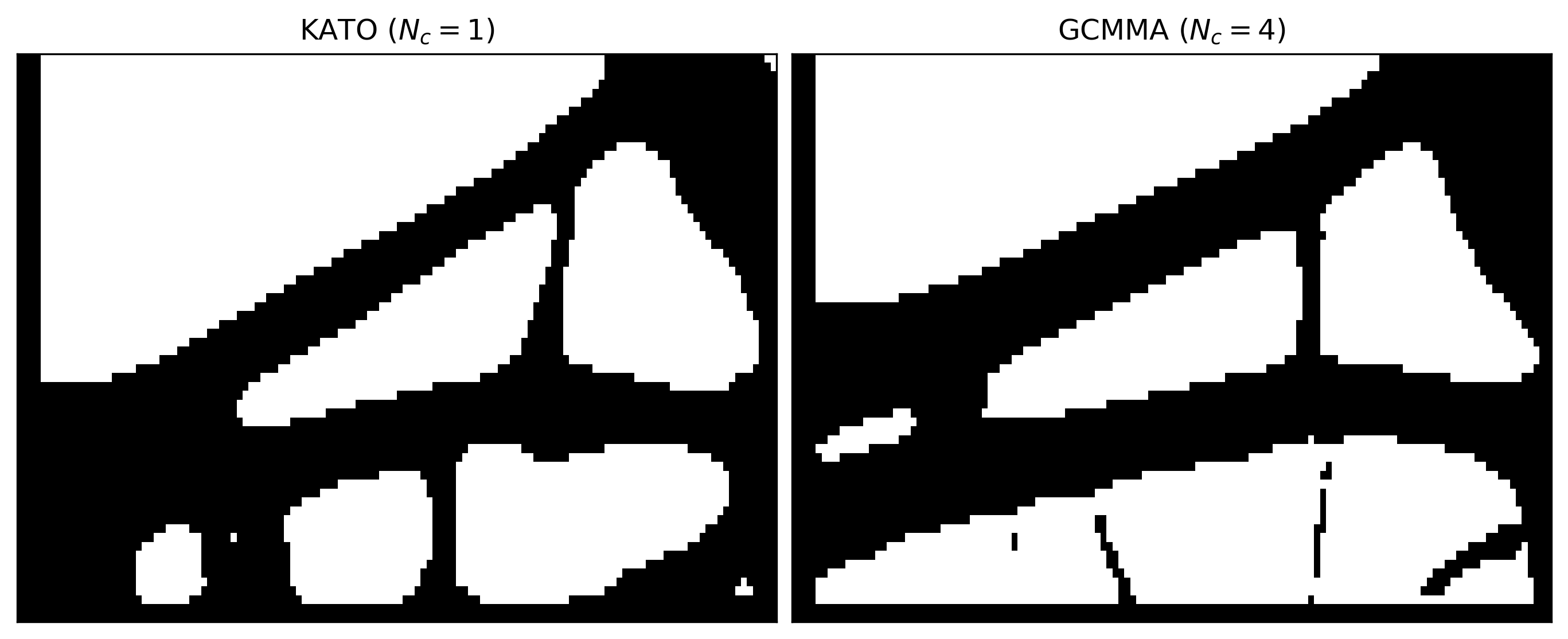}
  \caption{Optimized designs obtained from the Unrestricted KATO and GCMMA runs
    at 300~Hz. The labels report the connected-component count $N_c$.}
  \label{fig:frame_300hz_binary}
\end{figure}

Figure~\ref{fig:frame_300hz_binary} and Table~\ref{tab:frame_300hz} compare the
two Unrestricted solutions obtained at 300~Hz. Relative to the uniform starting
field (172.02~dB at 300~Hz), KATO and GCMMA reduce the binary AIP to 139.76 and
139.30~dB, respectively. The KATO binary has 59.4$\times$ lower static
compliance, a lower pre-threshold non-discreteness measure ($M_{nd}=8.0\%$
versus $32.9\%$), and one connected component. The GCMMA binary has four
components: its principal component contains 99.57\% of the solid elements, and
three islands contain the remaining 16 elements. The connected KATO result
is consistent with the architectural prior and spatial regularization of the
neural generator.

\begin{table}[pos=h]
  \centering
  \caption{Performance of the Unrestricted KATO and GCMMA frame designs
    optimized at 300~Hz. The non-discreteness measure $M_{nd}$ uses the
    continuous final fields; AIP, compliance, stress, and $N_c$ use the
    post-processed designs, and the FRF peaks are obtained from the 1--500~Hz
    sweeps.}
  \label{tab:frame_300hz}
  \footnotesize
  \setlength{\tabcolsep}{7pt}
  \begin{tabular}{@{}lccccccc@{}}
    \toprule
    Method & \makecell{AIP at 300~Hz\\(dB)} & \makecell{Compliance\\(N\,m)} &
    \makecell{Stress $p$-norm\\(MPa)} & \makecell{$M_{nd}$\\(\%)} & $N_c$ &
    \makecell{FRF peak\\(Hz)} & \makecell{Peak AIP\\(dB)} \\
    \midrule
    KATO  & 139.76 & 3.74   & 40.95 & 8.0  & 1 & 425 & 146.67 \\
    GCMMA & 139.30 & 222.12 & 43.67 & 32.9 & 4 & 382 & 149.40 \\
    \bottomrule
  \end{tabular}
\end{table}

\begin{figure}[pos=h]
  \centering
  \includegraphics[width=0.88\textwidth]{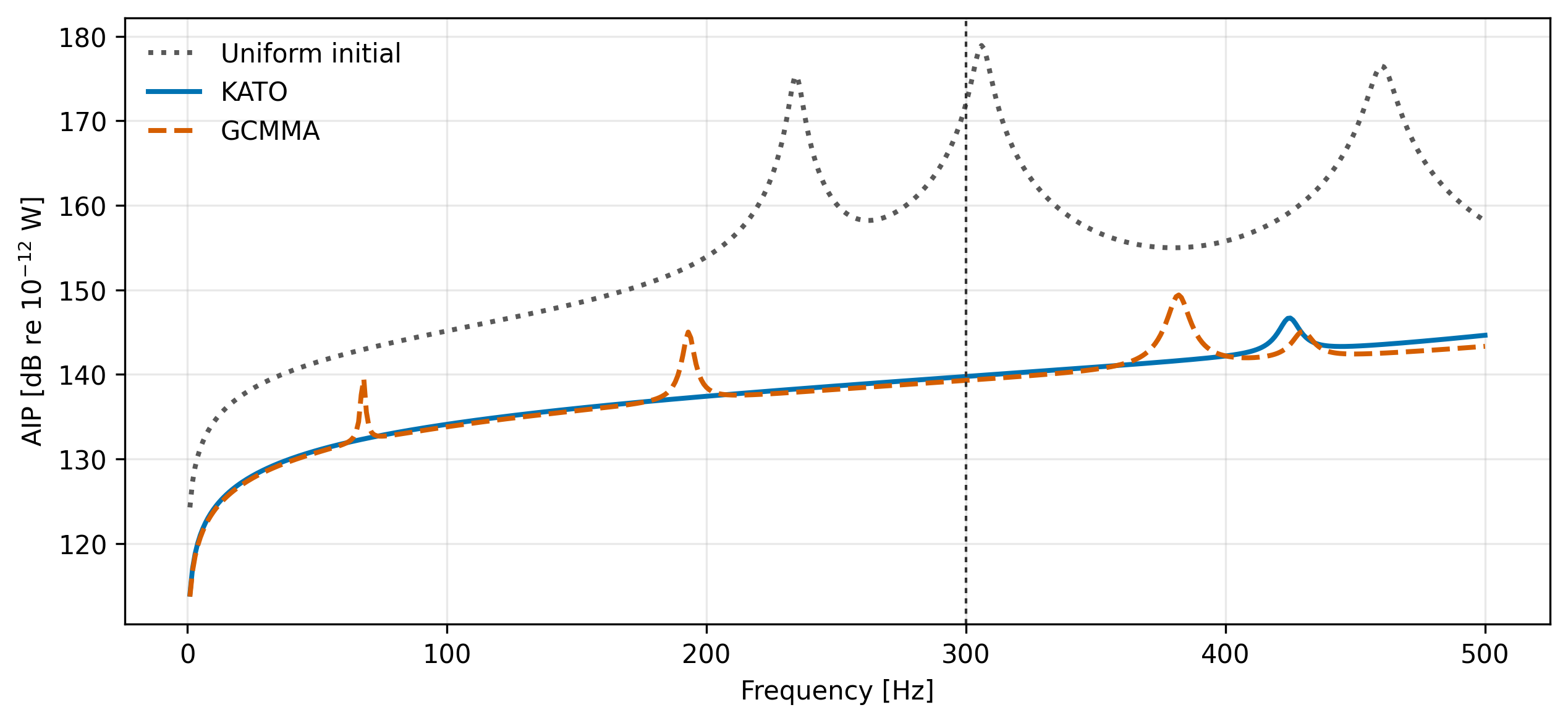}
  \caption{AIP frequency responses over 1--500~Hz for the uniform initial field
    and the optimized KATO and GCMMA designs at 300~Hz. The vertical dashed line
    marks the design frequency.}
  \label{fig:frame_300hz_frf}
\end{figure}

Figure~\ref{fig:frame_300hz_frf} compares the 1--500~Hz responses of the uniform
starting field and the two binary designs. The swept response in Figure~\ref{fig:frame_300hz_frf} provides a stricter check
than the design-frequency value alone. Optimization moves the dominant
load-participating peak away from the 306~Hz peak of the uniform field. The KATO
peak occurs at 425~Hz and is 2.73~dB below the GCMMA peak at 382~Hz. Thus, in
this near-resonant test, KATO gives a slightly higher AIP at the selected
frequency but a lower worst-case AIP over 1--500~Hz, together with markedly
higher static stiffness and a connected binary topology.

\section{Discussion}\label{sec:discussion}

\subsection{Topology regularization, stress, and binary performance}

Binary re-analysis evaluates the thresholded 0--1 layouts, while $M_{nd}$ characterizes non-discreteness in the continuous density fields
\cite{sigmund2013topology, vatanabe2016topology}. In the frame
cases, thresholding removes gray bridges from several GCMMA fields, whereas the
spatially correlated KATO parameterization retains coherent load paths. The
lower $M_{nd}$ and connected KATO binaries support the role of the generator as
an architectural prior and implicit regularizer.

The aggregated stress treatment lowers the dynamic stress $p$-norm in both deck
families and in the frame, with small changes in AIP. KATO incorporates this
trade-off through objective regularization, while GCMMA uses an explicit
constraint. Local stress constraints and stress-relaxed interpolation provide
possible extensions \cite{cheng1997varepsilon, le2010stress,
holmberg2013stress, bruggi2008mixed}.

Helmholtz filtering and stress regularization act through complementary
mechanisms. The Helmholtz PDE filter suppresses density variation over its
physical radius, while the dilated, intermediate, and eroded projections define
the retained solid and void geometry across the threshold interval
\cite{lazarov2011filters, wang2011projection}. This combination controls solid
and void feature sizes and preserves the principal load paths under erosion and
dilation; the resulting deck layouts contain fewer, thicker internal members.

The stress term addresses a different design requirement by discouraging highly
stressed load paths. The frame results separate these roles clearly: Helmholtz
filtering controls feature size while preserving AIP performance, whereas
stress regularization produces the main reduction in the stress $p$-norm.

\subsection{Relation to prior AIP topology optimization}

This work minimizes structural AIP rather than acoustic radiated power
directly. Under steady-state harmonic excitation, the time-averaged active
power supplied to a passive coupled structural--acoustic system is balanced by
internal dissipation and energy transmitted from the structure, including
acoustic radiation \cite{fahy2007sound, ross1987mechanics}. Minimizing AIP
therefore reduces the mechanical energy available for transmission and acoustic
radiation during early structural design. Coupled structural--acoustic analysis
can subsequently resolve radiation efficiency and fluid-loading effects
\cite{du2007minimization, neofytou2025automatic}.

Silva et al.\ \cite{silva2020topology} established the viability of AIP
as a forced-vibration objective using SIMP on a 2D cantilever benchmark
($\approx 40\,000$ free DOFs). This work applies AIP-based TO to ship-deck and foundation
structures with physical units, feature-size control, stress regularization,
and binary re-analysis. The convolutional generator adds an architectural prior
to these explicit controls and preserves connected load paths in the reported
designs.

\subsection{Computational performance}

The experiments were executed on a workstation equipped with an Intel
Core~i7-14700F CPU (20 cores, up to 5.40~GHz) and 32~GB of dual-channel
DDR5-5600 memory.

The KATO and GCMMA runs have comparable cost for the two variants without stress
constraints, while KATO is 6.4--10.4$\times$ faster for the implemented stress-aware formulations
(Table~\ref{tab:frame_comparison}). This timing comparison retains the characteristic stress treatment of each method: KATO uses objective penalization, whereas the conventional density-based GCMMA benchmark uses an explicit stress constraint. The additional GCMMA cost comes from its
inner constrained subproblems.

The forward-and-adjoint evaluation was also profiled with three solver backends
(Table~\ref{tab:backend_cost}). Replacing SciPy SuperLU with uncached PARDISO
reduces the frame and deck step times by 2.7$\times$ and 5.3$\times$,
respectively. Reusing the PARDISO symbolic factorization and real-block sparsity
structure gives 11.0$\times$ and 26.8$\times$ speedups relative to SuperLU, with
gradient differences at $10^{-11}$ or below.

\begin{table}[pos=h]
  \centering
  \caption{Sparse-solver timing for one forward-and-adjoint evaluation of the
    18~Hz frame and 100~Hz deck cases. All backends use the same objective and
    adjoint path; speedups and gradient errors are measured relative to
    SuperLU.}
  \label{tab:backend_cost}
  \footnotesize
  \setlength{\tabcolsep}{4pt}
  \begin{tabular}{@{}llccc@{}}
    \toprule
    Case & Backend & \makecell{$t_{\mathrm{step}}$\\(ms)} &
    \makecell{Speedup\\vs. SuperLU} & \makecell{Gradient\\rel. err.} \\
    \midrule
    Frame 18~Hz & SuperLU & 919 & $1.0\times$ & 0 \\
                & PARDISO stock & 347 & $2.7\times$ & $8.5\times10^{-13}$ \\
                & PARDISO cached & 84 & $11.0\times$ & $7.7\times10^{-13}$ \\
    Deck 100~Hz & SuperLU & $4\,902$ & $1.0\times$ & 0 \\
                & PARDISO stock & 929 & $5.3\times$ & $2.8\times10^{-11}$ \\
                & PARDISO cached & 183 & $26.8\times$ & $2.9\times10^{-11}$ \\
    \bottomrule
  \end{tabular}
\end{table}

The 200-mode basis requires approximately 400--2900~s per deck structure, but it
reduces each frequency point to a small dense solve. At the full-order
spot-checks, the corrected ROM is 990--2125$\times$ faster
(Table~\ref{tab:rom_validation}). The basis cost is amortized over the frequency
sweep.

Table~\ref{tab:rom_validation} quantifies the accuracy and cost for the double-skin
unrestricted deck panel: the residual-flexibility correction lowers the active-power
error from up to $11.8\%$ (uncorrected) to below $0.02\%$ across the five
50--400~Hz full-order spot checks. The correction accounts for the
1.7--16.7\% residual-flexibility fraction across the ten layouts.

\begin{table}[pos=h]
  \centering
  \caption{Accuracy and per-frequency cost of the uncorrected and
    residual-flexibility-corrected 200-mode ROM for the double-skin Unrestricted
    design at five full-order spot-check frequencies. The AIP error
    $\epsilon_\Pi$ follows Eq.~\eqref{eq:rom_error}; FOM and ROM times are from
    the dedicated PARDISO benchmark.}
  \label{tab:rom_validation}
  \footnotesize
  \setlength{\tabcolsep}{4pt}
  \begin{tabular}{@{}ccccccc@{}}
    \toprule
    \makecell{$f$\\(Hz)} & \makecell{AIP$_{\mathrm{FOM}}$\\(dB)} &
    \makecell{$\epsilon_\Pi$ uncorr.\\(\%)} & \makecell{$\epsilon_\Pi$ corr.\\(\%)} &
    \makecell{$t_{\mathrm{FOM}}$\\(s)} & \makecell{$t_{\mathrm{ROM}}$\\(s)} &
    Speedup \\
    \midrule
    50  & $102.33$ & 11.75 & 0.004 & 742  & 0.65 & $1140\times$ \\
    100 & $105.38$ & 11.65 & 0.001 & 661  & 0.67 & $990\times$  \\
    150 & $107.21$ & 11.47 & 0.003 & 662  & 0.63 & $1050\times$ \\
    250 & $109.85$ & 10.43 & 0.015 & 1011 & 0.69 & $1470\times$ \\
    400 & $117.07$ & 3.17  & 0.014 & 1316 & 0.62 & $2125\times$ \\
    \bottomrule
  \end{tabular}
\end{table}

\subsection{Limitations}

\begin{enumerate}
  \item \textbf{Material and excitation scope.} The study considers isotropic
    steel; the deck optimizations target 100~Hz and the primary
    frame study targets 18~Hz, with one additional matched comparison at 300~Hz.
    Multi-material formulations \cite{birman2018review, zenkert1997handbook}
    and multi-frequency objectives \cite{olhoff2016generalized} would extend
    the applicability. The swept-frequency post-evaluation confirms that TO
    layouts retain lower AIP near the design frequency, but the X-core reference
    becomes competitive at selected higher-frequency peaks.
  \item \textbf{Structural AIP scope.} The optimization targets structural
    AIP. Direct radiated sound power optimization requires a coupled
    structural-acoustic model (e.g., BEM \cite{junger1986sound}), which is the
    natural next step toward full URN minimization.
  \item \textbf{Manufacturing scope.} Helmholtz filtering
    \cite{lazarov2011filters} and Heaviside projection
    \cite{wang2011projection} provide mesh-independent feature-size control
    and promote discrete designs, but they do not impose
    fabrication-specific constraints such as weld accessibility and direction,
    plate rolling direction, joint detailing, or assembly requirements
    \cite{vatanabe2016topology, jihong2021review}. Incorporating these constraints
    is a future direction for production-oriented ship structural design.
  \item \textbf{2D optimization with 3D re-analysis.} The deck topology is
    optimized in 2D and evaluated in finite-depth 3D models. Full 3D TO would
    permit through-thickness material variation
    \cite{aage2017giga, liu2014efficient}.
\end{enumerate}

\section{Conclusions}\label{sec:conclusion}

Active input power minimization under harmonic excitation was applied to
two ship-related structural design problems using the KATO neural
reparameterized topology optimization framework with convolutional
Kolmogorov--Arnold network (cKAN) generators. The resulting binary layouts
produced by KATO remain structurally connected in the final evaluations and are
compatible with feature-size control and stress regularization.

The main findings are:
\begin{enumerate}
  \item AIP provides a non-negative forced-vibration objective for frequency-domain ship TO and avoids the sign ambiguity of dynamic
    compliance near resonances and antiresonances. Both
    deck TO families reduce AIP below the engineering references (stiffened panel and X-core sandwich panel) in the 2D cross-sectional models at 100~Hz, and
    the advantage is retained in the extruded 3D models.
  \item Stress regularization lowers the dynamic stress $p$-norm with small AIP
    changes. For the deck variants, Helmholtz filtering and three-field
    projection control solid and void feature sizes and preserve the main load
    paths after binarization.
  \item For the foundation frame at 18~Hz, KATO matches GCMMA in AIP within
    0.5~dB for three variants while giving 22--36$\times$ lower binary
    compliance. At 300~Hz, KATO gives 59.4$\times$ lower binary compliance and a 2.73~dB lower maximum
    AIP over 1--500~Hz. Its binary layouts remain connected in both comparisons,
    consistent with the generator's architectural prior.
  \item Cached PARDISO reduces the forward-and-adjoint cost by up to
    26.8$\times$ relative to SuperLU. The corrected 200-mode model reproduces
    the full-order spot checks within 0.02\% and evaluates each frequency point
    990--2125$\times$ faster.
\end{enumerate}

Future work could integrate structural-acoustic coupling for direct
underwater radiated noise optimization and extend the framework to broadband
random excitation.

\section*{Statements and Declarations}\subsection*{Funding}This research was financially supported by the Natural Sciences and Engineering Research Council of Canada (NSERC) through Discovery Grant RGPIN-2025-04421 and Alliance Advantage Grant ALLRP 607677-25. Additional financial support was provided by Seaspan Shipyards.

Formal analysis, Investigation, Writing~-- original draft.
Writing~-- review \& editing.

\subsection*{Data availability}The data and code supporting the findings of this study are available from the corresponding author upon reasonable request. They will be released at \url{https://github.com/ysyysy115/KATO_aip} after the manuscript has been accepted in a journal.

\section*{Acknowledgements}

The authors thank Norbert Schumacher, P.Eng., of Robert
Allan Ltd. for his constructive comments on ship structural optimization.



\bibliographystyle{model1-num-names}
\setlength{\bibsep}{0pt}
\renewcommand{\bibfont}{\fontsize{7.8}{8.8}\selectfont}
\bibliography{cas-refs}

\end{document}